\documentclass[11pt]{article}

\usepackage{amsmath}
\usepackage{amsthm}
\usepackage{amssymb}
\usepackage{algorithm}
\usepackage{algpseudocode}
\usepackage{authblk}
\usepackage{bbm}
\usepackage{cite} 
\usepackage{color}
\usepackage{comment}
\usepackage{enumitem}
\usepackage{float}
\usepackage[T1]{fontenc}
\usepackage[margin=1in]{geometry}
\usepackage{graphicx}
\usepackage{hyperref}
\usepackage{mathtools}
\usepackage{subfigure}
\usepackage{tikz}
\usepackage{url}
\usepackage{tablefootnote}

\newtheorem{theorem}{Theorem}[section]

\newtheorem{assumption}[theorem]{Assumption}

\newcommand{\poly}[1]{\operatorname{poly} \lrb{#1}}
\renewcommand{\log}[1]{\operatorname{log} \lrb{#1}}
\newcommand{\logb}[1]{\operatorname{log}_2 \lrb{#1}}

\newcommand{\lrb}[1]{\left ( #1 \right )}
\newcommand{\lrsb}[1]{\left [ #1 \right ]}
\newcommand{\lrcb}[1]{\left \{ #1 \right \}}

\newcommand{\abs}[1]{\left | #1 \right |}
\newcommand{\norm}[1]{\left \| #1 \right \|}
\newcommand{\myspan}[1]{\operatorname{span}\lrb{#1}}
\newcommand{\mytr}[1]{\operatorname{tr}\lrb{#1}}
\newcommand{\supp}[1]{\operatorname{supp}\lrb{#1}}

\newcommand{\ket}[1]{\left| #1 \right\rangle}
\newcommand{\bra}[1]{\left\langle #1 \right|}
\newcommand{\ketbra}[2]{|#1\rangle \langle #2|}
\newcommand{\braket}[2]{\langle #1|#2\rangle }

\newcommand{\zon}{\{0,1\}^n}
\newcommand{\zo}{\{0,1\}}

\newcommand{\R}{\mathbb{R}}

\newcommand{\Z}{\mathbb{Z}}
\renewcommand{\P}[1]{\mathbb{P}\lrsb{#1}}
\newcommand{\E}[1]{\mathbb{E}\lrsb{#1}}

\newcommand{\defeq}{\coloneqq}
\renewcommand{\emptyset}{\varnothing}
\newcommand{\cvar}[1]{ \operatorname{CVaR}_{\alpha}\lrb{#1}}

\newcommand{\argmin}{\operatorname{argmin}}

\newcommand{\indicator}[1]{\mathbbm{1}_{#1}}

\newcommand{\qaoaz}{\overline{\mathrm{QAOA}}}
\DeclareMathOperator{\gmqaoa}{GM-QAOA}

\DeclareMathOperator{\kz}{KZ}
\DeclareMathOperator{\fl}{FL}

\DeclareMathOperator{\cbqoa}{CBQOA}

\DeclareMathOperator{\pogs}{POGS}

\title{Classically-Boosted Quantum Optimization Algorithm}

\author{Guoming Wang\thanks{\texttt{guoming.wang@zapatacomputing.com}. Zapata Computing Canada Inc.}}

\begin{document}

\maketitle

\begin{abstract}
Considerable effort has been made recently in the development of heuristic quantum algorithms for solving combinatorial optimization problems. Meanwhile, these problems have been studied extensively in classical computing for decades. In this paper, we explore a natural approach to leveraging existing classical techniques to enhance quantum optimization. Specifically, we run a classical algorithm to find an approximate solution and then use a quantum circuit to search its ``neighborhood'' for higher-quality solutions. We propose the Classically-Boosted Quantum Optimization Algorithm (CBQOA) that is based on this idea and can solve a wide range of combinatorial optimization problems, including all unconstrained problems and many important constrained problems such as Max Bisection, Maximum Independent Set, Minimum Vertex Cover, Portfolio Optimization, Traveling Salesperson and so on. A crucial component of this algorithm is an efficiently-implementable continuous-time quantum walk (CTQW) on a properly-constructed graph that connects the feasible solutions. CBQOA utilizes this CTQW and the output of an efficient classical procedure to create a suitable superposition of the feasible solutions which is then processed in certain way. This algorithm has the merits that it solves constrained problems without modifying their cost functions, confines the evolution of the quantum state to the feasible subspace, and does not rely on efficient indexing of the feasible solutions. We demonstrate the applications of CBQOA to Max 3SAT and Max Bisection, and provide empirical evidence that it outperforms previous approaches on these problems.
\end{abstract}

\section{Introduction}

In recent years, there has been growing interest in utilizing near-term quantum devices to solve challenging problems in combinatorial optimization. In their seminal paper \cite{farhi2014quantum}, Farhi et al. introduced the Quantum Approximate Optimization Algorithm (QAOA) which is inspired by Trotterization of adiabatic quantum computing \cite{farhi2000quantum}. The QAOA circuit starts with the uniform superposition of all possible solutions, and alternately applies the time evolutions of the cost Hamiltonian and a mixing Hamiltonian (which are called the phase separators and mixing operators, respectively) multiple times, before measuring the final state in the computational basis. A classical optimizer is employed to tune the parameters (i.e. the evolution times) to maximize the probability of obtaining high-quality solutions from the final state. Numerous studies have been carried out to understand the characteristics of QAOA (e.g. \cite{farhi2016quantum, brandao2018fixed, lloyd2018quantum, mcclean2018barren, guerreschi2019qaoa, akshay2020reachability, zhou2020quantum, dalzell2020many, farhi2020quantum1, farhi2020quantum2, harrigan2021quantum, mcclean2021low}) and to propose techniques to improve its performance (e.g. \cite{wecker2016training, hadfield2019quantum, gilyen2019optimizing, shaydulin2019multistart, verdon2019learning, bravyi2020obstacles, stokes2020quantum, wauters2020reinforcement, khairy2020learning, barkoutsos2020improving, wierichs2020avoiding, egger2021warm, bartschi2020grover}).

Meanwhile, combinatorial optimization has been extensively studied in classical computing for decades, and many sophisticated techniques (e.g. SDP relaxation and spectral methods) have been developed to tackle these problems. It is conceivable that one could leverage these techniques to enhance quantum optimization without incurring much overhead. Surprisingly, this has been rarely done in previous works (except that classical techniques for continuous optimization have been used to tune the parameters in the ansatz circuits). A few exceptions are the recent works on \emph{warm-starting} quantum optimization \cite{egger2021warm, tate2020bridging, tate2021classically, van2021quantum}. In particular, Egger et al. \cite{egger2021warm} proposed the warm-started QAOA (WS-QAOA) in which the initial state and mixing operator are constructed based on the continuous solution of the Quadratic Programming (QP) or Semidefinite Programming (SDP) relaxation of the original problem, and showed that this algorithm outperforms standard QAOA at low depths. The reason for the superiority of this algorithm can be understood intuitively as follows. In standard QAOA, the initial state is the equal superposition of all  possible solutions regardless of their qualities. On the other hand, in WS-QAOA, the initial state is a non-uniform superposition of the possible solutions such that a discrete solution has large amplitude if and only if it is close to the continuous solution of QP or SDP relaxation -- which itself resembles the optimal discrete solution in some way. As a consequence, the high-quality solutions have larger amplitudes in the initial state of WS-QAOA than in the initial state of standard QAOA. Thus, WS-QAOA requires fewer layers to amplify them to sufficiently large numbers. This example demonstrates that one can utilize prior information provided by efficient classical procedures to improve quantum optimization algorithms.

To date, most works on QAOA and its variants have focused on solving unconstrained optimization problems (especially Max Cut). In practice, however, we often encounter constrained optimization problems in which a variable assignment is a feasible solution if and only if it satisfies certain constraints on the variables. Traditionally, these problems are handled by adding a penalty term (which depends on the constraints) to the cost function and solving the modified problem by QAOA or its variants. However, this approach is less efficient than ideal, because it needs to search the whole space -- which could be much larger than the feasible subspace -- for a satisfactory solution. Furthermore, one needs to design the penalty term properly to ensure that the solution to the modified problem is a high-quality feasible solution to the original problem, which can be tricky. 

To better tackle constrained optimization problems, Hadfield et al. \cite{hadfield2019quantum} extended the original QAOA to the Quantum Alternating Operator Ansatz ($\qaoaz$) framework. The $\qaoaz$ circuit also alternates between the phase separators and mixing operators as in QAOA, but it starts with a suitable superposition of the \emph{feasible} solutions, and each mixing operator in this circuit is a unitary operator that satisfies two conditions: 1. It preserves the feasible subspace; 2. It provides the transition between all pair of states corresponding to feasible solutions. The authors explicitly constructed a variety of mixing operators (e.g. the XY mixers and permutation mixers) for many optimization problems. By design, the evolution of the quantum state in $\qaoaz$ is confined to the feasible subspace, and hence we always obtain a feasible solution from the measurement on the final state. This framework has been applied to various constrained problems and shown advantages over previous approaches \cite{hadfield2019quantum, wang2020xy, hodson2019portfolio, hodson2020finding, cook2020quantum}. 

Independently, Marsh and Wang \cite{marsh2019quantum, marsh2020combinatorial} proposed the Quantum-Walk-assisted Optimization Algorithm (QWOA) to handle constrained problems. QWOA shares some basic ideas with $\qaoaz$. In fact, the QWOA circuit can be viewed as a special instantiation of the $\qaoaz$ circuit in which each mixing operator is a continuous-time quantum walk (CTQW) \cite{farhi1998quantum} on a graph whose nodes correspond to the feasible solutions. To facilitate the implementation of the CTQWs, this algorithm requires that the feasible solutions can be efficiently indexed. Namely, assuming there are $M$ feasible solutions and they are sorted in a reasonable way, we can compute the $j$-th feasible solution efficiently for given any $j \in \lrcb{1,2,\dots,M}$. Then with the help of this indexing algorithm, we can efficiently implement the CTQWs on certain graphs (e.g. circulant and complete graphs) that connect the feasible solutions. As in $\qaoaz$, the evolution of the quantum state in QWOA is confined to the feasible subspace, and only feasible solutions are obtained from measuring the final state. This algorithm has been utilized to solve several constrained problems \cite{marsh2019quantum, marsh2020combinatorial, slate2021quantum}. 

We remark that for some constrained problems, such as Maximum Independent Set and Minimum Vertex Cover, efficient indexing of the feasible solutions might be difficult or even impossible. In fact, for such problems, it is hard to simply count the number of feasible solutions. Nevertheless, in these cases, it is possible to \emph{sample} the feasible solutions in a Markov Chain Monte Carlo (MCMC) fashion, where each step slightly modifies the current solution. This prompts us to use a different method to create a suitable superposition of the feasible solutions.

Finally, B\"{a}rtschi and Eidenbenz \cite{bartschi2020grover} developed the Grover-Mixer Quantum Alternating Operator Ansatz (GM-QAOA) to address constrained problems. The GM-QAOA circuit resembles that of standard QAOA, except that it starts with an equal superposition of all \emph{feasible} solutions, and the mixing operators are generalized reflections about this state. This algorithm can be also viewed as a special instantiation of QWOA in which the graph for the CTQW is the complete graph. It has the merit that the solutions with the same objective value are sampled with the same probability. However, the preparation of its initial state, i.e. the uniform superposition of all feasible solutions, could be difficult for some problems. Furthermore, this choice of the initial state is not ideal in many cases, as suggested by the numerical evidence in Section \ref{sec:experiments}.

In this paper, we present a hybrid quantum-classical algorithm named the \emph{Classically-Boosted Quantum Optimization Algorithm (CBQOA)} for solving a broad class of combinatorial optimization problems, including all unconstrained problems and many important constrained problems such as Max Bisection, Maximum Independent Set, Minimum Vertex Cover, Portfolio Optimization, Traveling Salesperson and so on. Our basic idea is quite straightforward: We run a classical algorithm to find an approximate solution (which is called the \emph{seed}) and then use a quantum circuit to search its ``neighborhood" for higher-quality solutions. Specifically, we show that as long as the domain of the problem satisfies certain conditions, one can construct a weighted undirected graph connecting the feasible solutions such that the CTQWs on this graph can efficiently implemented. Under this assumption, CBQOA calls an efficient classical procedure to generate a seed and runs the aforementioned CTQW starting at the seed to create a suitable superposition of the feasible solutions (which is the initial state of CBQOA). Then CBQOA amplifies the amplitudes of the high-quality solutions within this state by alternately applying the generalized reflections about the initial state and the time evolutions of the cost Hamiltonian multiple times, before measuring the final state in the computational basis. Our design of the ansatz circuit ensures that only feasible solutions are obtained from this measurement. As in other variational quantum algorithms, CBQOA employs an iterative optimizer to tune the circuit parameters to maximize the probability of receiving high-quality solutions from the final state. 

Table \ref{tab:comparison_works} compares the characteristics of CBQOA and the previous algorithms including standard QAOA, WS-QAOA, $\qaoaz$, QWOA and GM-QAOA. One can see that CBQOA is the only algorithm that utilizes efficient classical pre-processing to adaptively construct the ansatz circuit and initial state, solves constrained problems without modifying their cost functions, confines the evolution of the quantum state to the feasible subspace, and does not rely on efficient indexing of the feasible solutions. In other words, CBQOA possesses all the desired qualities simultaneously.

\begin{table*}[ht]
    \centering
    \begin{tabular}{cccccc}
          & Adaptive & Adaptive & Directly solves &  Evolution  & Relies on     \\
        Algorithm  & ansatz & initial & constrained &  confined to  & efficient    \\       
          & circuit? & state? &  problems? & feasible subspace? & indexing?  \\                   
         \hline
         QAOA & N & N & N & N & N \\
         WS-QAOA & Y & Y & N & N & N  \\
         $\qaoaz$ & N & Y  & Y & Y& N  \\
         QWOA & N & Y  & Y & Y & Y  \\
         GM-QAOA & N & N & Y & Y &N  \\
         CBQOA & Y & Y & Y & Y & N            
    \end{tabular}
    \caption{This table compares the characteristics of standard QAOA, WS-QAOA, $\qaoaz$, QWOA, GM-QAOA and CBQOA for combinatorial optimization, including whether the algorithm utilizes efficient classical pre-processing to adaptively construct the ansatz circuit and initial state, whether it solves constrained problems without modifying their cost functions, whether the evolution of the quantum state is confined to the feasible subspace, and whether it relies on efficient indexing of the feasible solutions. We remark that although $\qaoaz$ and QWOA allow for adaptive construction of the initial state in principle, this strategy has been rarely adopted in previous works. In addition, although GM-QAOA does not require efficient indexing of the feasible solutions per se, we are not aware of any problem for which the equal superposition of the feasible solutions can be quickly prepared but these solutions cannot be efficiently indexed.}
    \label{tab:comparison_works} 
\end{table*}

We empirically evaluate the performance of CBQOA on two combinatorial optimization problems -- Max 3SAT which is unconstrained and Max Bisection which is constrained -- and find that CBQOA outperforms well-known classical algorithms and GM-QAOA on these problems. 

The remainder of this paper is organized as follows. In Section \ref{sec:overview}, we give a high-level overview of the CBQOA algorithm. In Section \ref{sec:ctqw}, we describe how to design and implement a crucial component of CBQOA -- a CTQW on a properly-constructed graph that connects the feasible solutions. In Section \ref{sec:applications}, we demonstrate the applications of CBQOA to Max 3SAT and Max Bisection. In Section \ref{sec:experiments}, we empirically test the performance of CBQOA on these two problems and compare it with that of previous methods. Finally, Section \ref{sec:conclusion} concludes this paper.

\section{Overview of CBQOA}
\label{sec:overview}
Without loss of generality, any combinatorial optimization problem can be represented as a cost function $f: F \subseteq \zon \to \R$, where $F$ is the domain of the problem (i.e. the set of feasible solutions). The goal is to find a mimima of this function, i.e. $x^* = \argmin_{x \in F} f(x)$. To solve this problem on a quantum device, we encode $f$ into an $n$-qubit Ising Hamiltonian $H_f \defeq \sum_{x \in \zon} f(x) \ketbra{x}{x}$ (i.e. cost Hamiltonian) \footnote{We define $f(x)$ arbitrarily for each $x \in \zon \setminus F$.}, and assume that $e^{-i H_f \gamma}$ can be implemented with $\poly{n}$ elementary gates for arbitrary $\gamma \in \R$. This is true if $f$ can be written as an $O(\log{n})$-degree polynomial with $\poly{n}$ terms in its variables, which means that $H_f$ can be decomposed into $\poly{n}$ commuting $O(\log{n})$-local terms. This condition is satisfied by many natural combinatorial optimization problems.

\begin{figure}[ht]
    \centering
    \includegraphics[width=\linewidth]{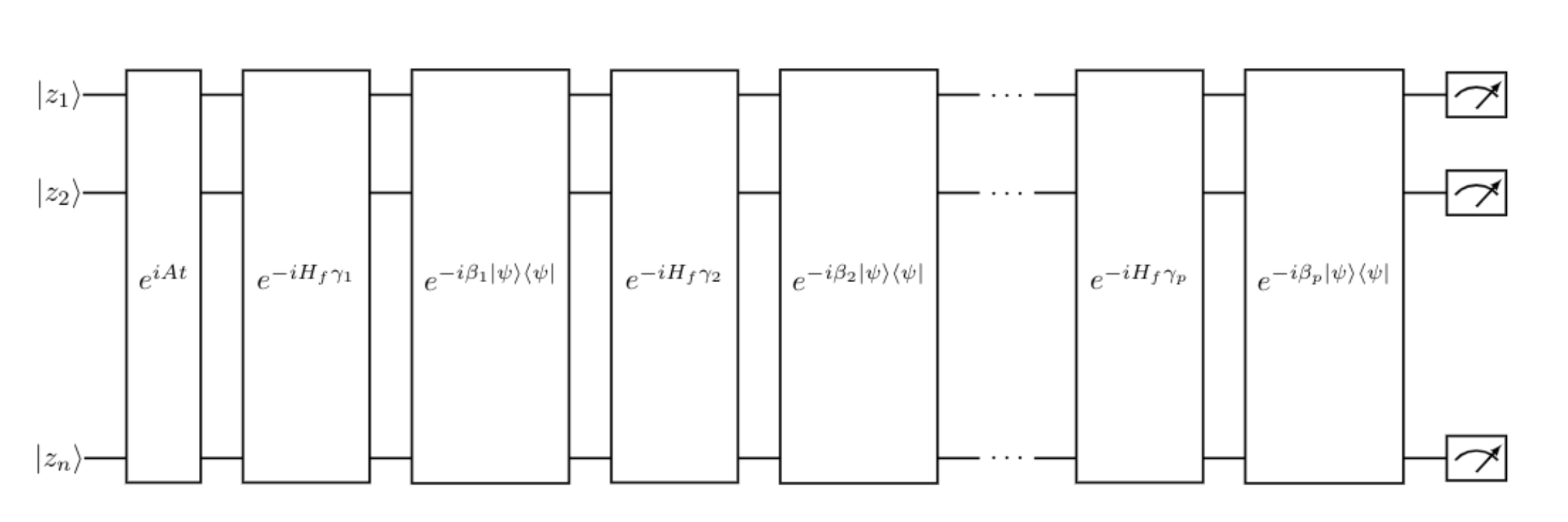}
    \includegraphics[width=0.7\linewidth]{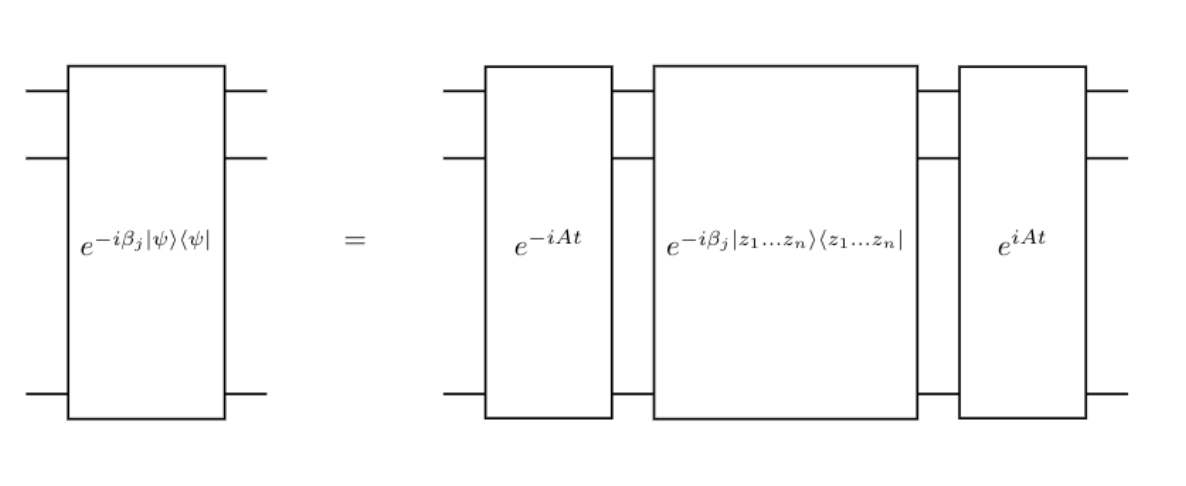}
    \caption{Quantum circuit for CBQOA. Here $z=(z_1,z_2,\dots,z_n)$ is a feasible solution generated by a classical algorithm, and $\ket{\psi}=e^{iAt} \ket{z}$ in which $A$ is defined in Section \ref{subsec:build_graph}.}
    \label{fig:cbqoa_circuit}
\end{figure}

Figure \ref{fig:cbqoa_circuit} illustrates the CBQOA circuit for tackling the above problem. It requires two elements to work:
\begin{itemize}
    \item An efficient classical algorithm for producing a feasible solution $z \in F$ (which will be referred to as the \emph{seed} from now on);
    \item An efficiently-implementable CTQW $e^{iAt}$ on a weighted undirected graph $G=(V, E, w)$ such that: (1) $G$ depends on $f$ and $z$; (2) $V=\zon$; (3) $F$ and $V \setminus F$ are disconnected in $G$; (4) The induced subgraph of $G$ on $F$ is the union of $\poly{n}$ connected components of $G$. 
\end{itemize}
We will describe the construction of $G$ and the implementation of $e^{iAt}$ in Section \ref{sec:ctqw}. By design, the state $\ket{\psi} \defeq e^{iAt} \ket{z}$ is a superposition of the feasible solutions, i.e. $\ket{\psi} \in \mathcal{H}_F \defeq \myspan{\ket{x}:~x \in F}$. Ideally, we want $z$'s neighborhood in $G$ to contain as many high-quality feasible solutions as possible, so that $e^{iAt} \ket{z}$ has large overlap with the states corresponding to those solutions for some small $t$. While this is difficult to achieve in general, we take advantage of the fact that for most optimization problems, similar solutions tend to have similar qualities, and use a well-performing classical approximation algorithm to generate the seed. This strategy proves to be effective in the experiments we conducted. 

After preparing the state $\ket{\psi}=e^{iAt} \ket{z}$, we alternately perform two types of unitary operations on it multiple times, receiving the ansatz state 
\begin{align}
\ket{\psi(\vec \beta, \vec \gamma)} \defeq e^{-i \beta_p \ketbra{\psi}{\psi}} e^{-i \gamma_p H_f}
e^{-i \beta_{p-1} \ketbra{\psi}{\psi}} e^{-i \gamma_{p-1} H_f}\dots 
e^{-i \beta_1 \ketbra{\psi}{\psi}} e^{-i \gamma_1 H_f} \ket{\psi},
\label{eq:ansatz_state_gm_qaoa}
\end{align}
where $p$ is the number of circuit layers, $\vec \beta=(\beta_1, \beta_2, \dots, \beta_p)$, $\vec \gamma = (\gamma_1, \gamma_2, \dots, \gamma_p) \in \R^p $ are tunable parameters. Following the convention, we call each $e^{-i \beta_j \ketbra{\psi}{\psi}}$ a \emph{mixing operator} and each $e^{-i \gamma_j H_f}$ a \emph{phase separator}, although our algorithm is arguably more similar to amplitude amplification \cite{brassard2002quantum} than to QAOA. Note that each mixing operator $e^{-i\beta_j \ketbra{\psi}{\psi}}$ can be implemented based on the equation:
\begin{align}
e^{-i\beta_j \ketbra{\psi}{\psi}}&=e^{iAt} e^{-i \beta_j \ketbra{z}{z}} e^{-iAt},
\end{align}
where $e^{-i \beta_j \ketbra{z}{z}}$ can be implemented with $O(n)$ elementary gates (with the help of $n-1$ ancilla qubits). So the CBQOA circuit is indeed efficient.

By construction, we have $\ket{\psi(\vec \beta, \vec \gamma)} \in \mathcal{H}_F$ for arbitrary $(\vec \beta, \vec \gamma) \in \R^{2p}$. So measuring it in the computational basis always yields a feasible solution. In fact, the evolution of the quantum state is confined to the feasible subspace $\mathcal{H}_F$ throughout the CBQOA circuit. This is beneficial when $F$ is much smaller than $\zon$, as it makes the search for high-quality feasible solutions more efficient. 

Following the suggestion of Ref. \cite{barkoutsos2020improving}, we tune the parameters $(\vec \beta, \vec \gamma)$ by minimizing the Conditional Value at Risk (CVaR) of the cost of a random solution sampled from $\ket{\psi(\vec \beta, \vec \gamma)}$. Specifically, the CVaR of a random variable $X$ for a confidence level $\alpha \in (0, 1]$ is defined as 
\begin{align}
\cvar{X} \defeq \E{X| X \le F_X^{-1}(\alpha)},
\end{align}
where $F_X$ is the cumulative density function of $X$. In other words, $\cvar{X}$ is the expected value of the lower $\alpha$-tail of the distribution of $X$. In our task, $X$ is defined as follows. When we measure $\ket{\psi(\vec \beta, \vec \gamma)}$ in the computational basis, we obtain a random solution $X_{\vec \beta, \vec \gamma}$ in $F$ such that 
\begin{align}
    \P{X_{\vec \beta, \vec \gamma}=x} = \abs{\braket{x}{\psi(\vec \beta, \vec \gamma)}}^2, \quad \forall x \in F.
    \label{eq:prob_dist_x}
\end{align}
Then we set $(\vec \beta, \vec \gamma)$ to be the solution to the following problem:
\begin{align}
    \min_{\vec \beta,\vec \gamma \in \R^p} \cvar{f(X_{\vec \beta, \vec \gamma})},
    \label{eq:qaoa_opt_prob}
\end{align}
where $\alpha \in (0, 1]$ is the confidence level. In particular, when $\alpha=1$, we have $\cvar{f(X_{\vec \beta, \vec \gamma})} = \E{f(X_{\vec \beta, \vec \gamma})}$. So in this case, we simply minimize the average cost of a random solution $X$ sampled from $\ket{\psi(\vec \beta, \vec \gamma)}$. The optimal choice of $\alpha$ depends on the specific situation (e.g. the problem to solve and its size) \footnote{In our experiments, we set $\alpha=0.5$ and find that this leads to satisfactory performance}.

Problem \ref{eq:qaoa_opt_prob} is solved by an iterative optimizer (e.g. ADAM or L-BFGS-B) in which each evaluation of $\cvar{f(X)}$ involves running the CBQOA circuit on a quantum device or simulating it on a classical computer. It turns out that the latter approach can be greatly accelerated by utilizing the technique in Appendix \ref{subsec:simulate_cbqoa_circuit}.

\section{State preparation by continuous-time quantum walks}
\label{sec:ctqw}
In this section, we describe how to design and implement the CTQW $e^{iAt}$ for state preparation in CBQOA. 

\subsection{Continuous-time quantum walks}
\label{subsec:ctqw}
A CTQW is the evolution of a quantum system under a Hamiltonian defined by the adjacency matrix of a graph. Formally, suppose $G=(V, E, w)$ is a weighted undirected graph, where $w: E \to \R$ assigns a weight to each edge. The continuous-time quantum walk $U(t)$ on $G$ at time $t$ is given by
\begin{align}
U(t)  \defeq e^{i At},
\end{align}
where $A$ is the adjacency matrix of $G$ (i.e. $A$ is a symmetric $|V| \times |V|$ matrix such that $A[u,v]=w_{u,v}$ if $(u,v)\in E$, and $0$ otherwise). The probability of a walk starting at vertex $v$ ending up at vertex $u$ at time $t$ is given by $\abs{\bra{u} U(t) \ket{v}}^2$. More generally, if we start the walk with a superposition of the vertices, i.e. $\ket{\psi} = \sum_{v \in V} \alpha_v \ket{v}$, the probability of the walk ending up at vertex $u$ at time $t$ is given by $\abs{\bra{u} U(t) \ket{\psi}}^2$. 

CTQWs possess different characteristics from continuous-time random walks (CTRWs). For example, the probability distribution of the vertices during a CTRW converges to a stationary distribution given enough time, but such a distribution does not exist for a generic CTQW. Namely, the probability distribution of the vertices during a generic CTQW constantly changes over time. Therefore, ones needs to carefully pick $t$ so that $U(t)\ket{\psi}$ has the desired property. In this work, we focus on the case where $t$ is small and hence $\abs{\bra{u}U(t)\ket{v}}$ is large only if $u$ is close to $v$. Namely, we would like to search $v$'s neighborhood for a vertex that meets certain conditions.

\subsection{Building the graph}
\label{subsec:build_graph}
For convenience, we introduce the following notation and definition. For any integer $n \ge 1$, let $[n]\defeq \lrcb{1,2,\dots,n}$. For any $S \subseteq [n]$, let $S^c \defeq [n] \setminus S$. For any $x \in \zon$, let $\supp{x} \defeq \lrcb{j \in [n]: x_j=1}$. Moreover, for any $S \subseteq [n]$ and $x \in \zon$, let $x_S$ be the restriction of $x$ to the coordinates in $S$, i.e. $x_S \defeq ( x_i:~i\in S)$. Finally, we say that a permutation $\tau: \zon \to \zon$ is \emph{$k$-local} if there exist $C, S \subseteq [n]$ and $y \in \zo^{|C|}$ such that: (1) $|S|=k$; (2) $C \cap S = \emptyset$; (3) $(\tau(x))_{S^c}=x_{S^c}$ for all $x \in \zon$; (4) If $C \neq \emptyset$, then $\tau(x)=x$ for all $x \in \zon$ satisfying $x_C \neq y$. 

We will build a weighted undirected graph $G=(V, E, w)$ such that $V=\zon$ (i.e. each vertex is a unique $n$-bit string), and $E$ and $w: E \to \R$ depend on the cost function $f: F\subseteq \zon \to \R$ and the seed $z \in F$, under the following assumption: 
\begin{assumption}
There exist $m=\poly{n}$ efficiently-computable permutations $\tau_1$, $\tau_2$, $\dots$, $\tau_m : \zon \to \zon$ such that:
\begin{enumerate}
\item For each $i \in [m]$, $\tau_i$ is $O(\log{n})$-local.
\label{locality}
\item For each $i \in [m]$, $\tau_i$ has order 2, i.e. $\tau_i(\tau_i(x)) = x$ for all $x \in \zon$, and $\tau_i$ is not the identity permutation.
\label{order}
\item There exist $k=\poly{n}$ subsets $F_1, F_2, \dots, F_k$ of $F$ such that:
\begin{enumerate}
    \item $(F_1, F_2, \dots, F_k)$ forms a partition of $F$; \label{partition}
    \item For each $i \in [m]$ and each $j \in [k]$, $\tau_i$ maps $F_j$ to itself, i.e. $F_j=\tau_i(F_j)\defeq \lrcb{\tau_i(x):~x \in F_j}$.\label{closure}
    \item For each $j \in [k]$, every two elements in $F_j$ can be transformed from one to another by a sequence of operations in $\mathcal{T}\defeq \lrcb{\tau_1,\tau_2,\dots,\tau_m}$. Namely, for arbitrary $x, y \in F_j$, $x \neq y$, there exist an integer $q \ge 1$ and $i_1, i_2, \dots, i_q \in [m]$ such that $y=\tau_{i_q}(\tau_{i_{q-1}}(\dots \tau_{i_1}(x)))$.
    \label{transitivity}
\end{enumerate} 
\end{enumerate}
\label{assumption}
\end{assumption}
This assumption holds for a wide range of combinatorial optimization problems, including all unconstrained problems (e.g. Max Cut, Max 3SAT), the constrained ones with permutation-invariant domains (e.g. Max Bisection), Maximum Independent Set, Minimum Vertex Cover, Portfolio Optimization, Traveling Salesperson, and so on \footnote{Assumption \ref{assumption} does not hold for some constrained problems, e.g. Knapsack. In this case, we needs convert the original problem to an unconstrained one by adding a penalty term (depending on the constraints) to the cost function, as done in previous works. Then we can apply our method to the modified problem. However, by doing so, we can no longer guarantee that only feasible solutions are obtained from measuring the ansatz state.}. Specifically, the permutations $\mathcal{T}= \lrcb{\tau_1,\tau_2,\dots,\tau_m}$ and the partition $(F_1, F_2, \dots, F_k)$ of the domain $F$ can be constructed for these problems as follows:
\begin{itemize}
    \item In an unconstrained problem, we have $F=\zon$ and can define $\mathcal{T}$ in multiple ways. The common choice is $\mathcal{T} \defeq \lrcb{\tau_1,\tau_2,\dots,\tau_n}$, where $\tau_i(x)=(x_1,\dots,x_{i-1}, \neg x_i, x_{i+1}, \dots, x_n)$ for all $x \in \zon$. Namely, $\tau_i$ flips the $i$-th bit of the input string. The corresponding partition of $F$ is the trivial partition $(F)$.
    
    \item In a constrained problem where $F$ is invariant under the permutations of the $n$ bits, i.e. $F =\sigma(F) \defeq \lrcb{(x_{\sigma(1)}, x_{\sigma(2)}, \dots, x_{\sigma(n)}):~x \in F}$ for all $\sigma \in S_n$, we define $\mathcal{T}$ as follows. Let $\tilde{G}=(\tilde{V}, \tilde{E})$ be any connected graph with vertex set $\tilde{V} = [n]$. Then let $\mathcal{T} \defeq \lrcb{\tau_{a,b}:~(a,b) \in \tilde{E}}$, where $\tau_{a,b}$ swaps the $a$-th and $b$-th bits of the input string. Namely, $(\tau_{a,b}(x))_a = x_b$,
    $(\tau_{a,b}(x))_b = x_a$, and $(\tau_{a,b}(x))_j = x_j$ for all $j \in [n]\setminus\lrcb{a, b}$, for all $x \in \zon$. The corresponding partition of $F$ is $(\bar{F}_{j_1}, \bar{F}_{j_2}, \dots, \bar{F}_{j_k})$ for some $0 \le j_1 < j_2 < \dots < j_k \le n$, where $\bar{F}_j \defeq \lrcb{x \in \zon: ~\sum_{i=1}^n x_i = j}$ for any $j \in \lrcb{0, 1, \dots, n}$.
    
    \item In the Maximum Independent Set problem, one is given a graph $G=(V, E)$ and needs to find a subset $W$ of $V$ such that no two vertices in $W$ are adjacent and $|W|$ is maximized. Suppose $|V|=n$. Then each subset of $V$ can be represented by an $n$-bit string in the natural way. We define $\mathcal{T} \defeq \lrcb{\tau_v:~v \in V}$, where $\tau_v$ satisfies that 
    \begin{align}
    (\tau_v(x))_v = \begin{cases}
    \neg x_v, & \textrm{if}~x_u=0,~\forall u \in N(v),\\
    x_v, & \textrm{otherwise}, 
    \end{cases}    
    \end{align}
    in which $N(v)$ denotes the set of $v$'s neighbors in $G$, and $(\tau_v(x))_u = x_u$ for all $u \in V \setminus \lrcb{v}$. The corresponding partition of $F$ is the trivial partition $(F)$. The Minimum Vertex Cover problem can be handled similarly, except that the roles of $1$ and $0$ are switched.

    \item In the Portfolio Optimization problem, one needs to minimize the cost function $f(s)=\lambda \sum_{i,j=1}^n \sigma_{i,j}s_is_j - (1-\lambda) \sum_{i=1}^n r_i s_i$ subject to the constraint $\sum_{i=1}^n s_i=A$, where $\sigma_{i,j} \in \R$, $r_i \ge 0$, $\lambda \in [0, 1]$ and $A \in [n]$ are given parameters, and $s_1,s_2,\dots,s_n \in \lrcb{1, -1, 0}$ are the variables \footnote{The meaning of the parameters and variables is as follows. There are $n$ assets under consideration. $r_i$ is the expected return of asset $i$, $\sigma_{i,j}$ is the covariance between the returns of assets $i$ and $j$, $A$ is the total net positions allowed within the portfolio, and $\lambda$ is the risk control parameter. $s_i=1, -1, 0$ represents long position, short position, no position on asset $i$, respectively.}. We can convert this problem into a problem with binary variables as follows. For each $i\in [n]$, we introduce two variables $x_{2i-1}, x_{2i} \in \zo$ such that $s_i=x_{2i-1} - x_{2i}$. Then the original domain is mapped to $F \defeq \lrcb{x \in \zo^{2n}:~\sum_{i=1}^n x_{2i-1} -\sum_{i=1}^n x_{2i} = A}$, and the cost function $f$ can be re-written in terms of $x_1,x_2,\dots,x_{2n}$. We define $\mathcal{T}$ as follows. Let $\tilde{G}_1=(\tilde{V}_1, \tilde{E}_1)$ and $\tilde{G}_2=(\tilde{V}_2, \tilde{E}_2)$ 
    be two arbitrary connected graphs with vertex sets $\tilde{V}_1 = \tilde{V}_2 = [n]$. Then let $\mathcal{T} \defeq \lrcb{\tau_{2i-1,2j-1}:~(i,j) \in \tilde{E}_1} \cup \lrcb{\tau_{2i,2j}:~(i,j) \in \tilde{E}_2}$, where $\tau_{a,b}$ swaps the $a$-th and $b$-th bits of the input string. Namely, $(\tau_{a,b}(x))_a = x_b$, $(\tau_{a,b}(x))_b = x_a$, and $(\tau_{a,b}(x))_j = x_j$ for all $j \in [2n]\setminus\lrcb{a, b}$, for all $x \in \zo^{2n}$. The corresponding partition of $F$ is $(\bar{F}_0, \bar{F}_1, \dots, \bar{F}_{n-A})$ where $\bar{F}_j \defeq \lrcb{x\in \zo^{2n}:~\sum_{i=1}^n x_{2i-1}=A+j, ~\sum_{i=1}^n x_{2i}=j}$ for each $j$.

    \item In the Traveling Salesperson problem, one is given a list of $n$ cities and the distances between each pair of cities, and needs to find the shortest route that visits each city exactly once and returns to the origin city. Each feasible solution is a permutation of $[n]$, and can be represented by an $n\lceil \logb{n}\rceil$-bit string. Namely, every feasible solution is represented by some $x=(\vec{x}_1, \vec{x}_2,\dots,\vec{x}_n)$, where $\vec{x}_i=(x_{i,1},x_{i,2},\dots,x_{i,t}) \in \zo^{t}$ indexes the $i$-th city on the route, in which $t=\lceil \logb{n}\rceil$. We call $\vec{x}_i$ the $i$-th segment of $x$. Then we define $\mathcal{T}$ as follows. Let $\tilde{G}=(\tilde{V}, \tilde{E})$ be any connected graph with vertex set $\tilde{V} = [n]$. Then let $\mathcal{T} \defeq \lrcb{\tau_{a,b}:~(a,b) \in \tilde{E}}$, where $\tau_{a,b}$ swaps the $a$-th and $b$-th segments of the input string. Namely, for arbitrary $x \in \zon$, we have $(\tau_{a,b}(x))_{a,j}=(\tau_{a,b}(x))_{b,j}$, for all $j \in [t]$; $(\tau_{a,b}(x))_{b,j}=(\tau_{a,b}(x))_{a,j}$, for all $j \in [t]$;    
    $(\tau_{a,b}(x))_{c,j}=(\tau_{a,b}(x))_{c,j}$, for all $j \in [t]$, for all $c \in [n]\setminus \lrcb{a,b}$. The corresponding partition of $F$ is the trivial partition $(F)$.

\end{itemize}
One can verify that all of the above constructions satisfy the conditions in Assumption \ref{assumption}, and can generalize them to similar combinatorial optimization problems. 

Now under Assumption \ref{assumption}, we build $G=(\zon, E, w)$ as follows. The edge set $E$ is the disjoint union of $m$ edge sets $E_1$, $E_2$, $\dots$, $E_m$, where $E_i=\lrcb{(x, \tau_i(x)):~x \in \zon,~x \neq \tau_i(x)}$ for $i=1,2,\dots,m$. Moreover, all the edges in  $E_i$ share the same weight $w_i \in (0, 1)$ (which will be determined later) for each $i \in [m]$. Formally, the adjacency matrix $A$ of this graph is defined as
\begin{align}
A = \sum_{i=1}^m w_i H_i,
\label{eq:def_a}
\end{align}
where $H_i$ is a matrix given by
\begin{align}
H_i = \sum_{x \in \zon} \indicator{\tau_i(x) \neq x} \ket{\tau_i(x)}\bra{x},
\label{eq:def_hi}
\end{align}
where $\indicator{}$ is the indicator function. By construction, $H$ has zero diagonal entries and non-negative off-diagonal entries. Moreover, by conditions \ref{order}, \ref{partition} and \ref{closure} of Assumption \ref{assumption}, we know that $H_i$ is Hermitian, has eigenvalues in $\lrcb{1, -1, 0}$, and satisfies
\begin{align}
H_i = \sum_{j=1}^k P_j H_i P_j + (I-P) H_i (I-P),    
\end{align}
where $P_j = \sum_{x \in F_j} \ket{x}\bra{x}$, for $j=1,2,\dots,k$, and $P=\sum_{j=1}^k P_j$. It follows that $A$ is a valid adjacency matrix and satisfies
\begin{align}
A = \sum_{j=1}^k P_j A P_j + (I-P) A (I-P),    
\end{align}
which means that $F_1$, $F_2$, $\dots$, $F_k$ and $V \setminus F$ are disconnected in $G$. Furthermore, conditions \ref{closure} and \ref{transitivity} of Assumption \ref{assumption} imply that for each $j \in [k]$, every two vertices in $F_j$ are connected by a path through  vertices in $F_j$ only. Thus, the induced subgraph of $G$ on $F_j$ is a connected component of $G$ for each $j \in [k]$.

It remains to determine the edge weights $w_i$'s. In principle, we want to assign a large weight $w_i$ to each edge in $E_i$ if $\tau_i(x)$ is likely to be a better solution than $x$ for a typical $x$ close to $z$, and a small weight otherwise. Meanwhile, we do not want to be extremely biased towards a particular permutation $\tau_i$, so the magnitude of $w_i$ needs to stay in a reasonable range. For these reasons, we set $w_i$ to be a sigmoid function of $\eta_i:=f(z) - f(\tau_i(z))$:
\begin{align}
w_i = w_i(\theta) \defeq \dfrac{1}{ 1 + e^{-\eta_i \theta}}=\dfrac{1}{ 1 + e^{-\theta \lrsb{f(z) - f(\tau_i(z))} }},
\end{align}
where $\theta \in \R$ is a tunable parameter. Note that we always have $w_i \in [0, 1]$, and $w_i$ is a monotonically increasing function of $\eta_i$ assuming $\theta>0$ is fixed. (In particular, setting $\theta=0$ leads to $w_i=1/2$ for all $i$'s, which means that all the edges in $G$ share the same weight. In this case, we treat all the permutations $\tau_i$'s equally.) We acknowledge that this choice of $w_i$ might be not optimal, and leave it as future work to find a better way to assign the edge weights.

\subsection{Implementing the CTQW}
\label{subsec:implement_ctqw}
In principle, since $A$ is a $2^n \times 2^n$ Hermitian matrix and each row of $A$ contains $\poly{n}$ non-zero entries whose locations and values can be efficiently computed, we could use various Hamiltonian simulation techniques (e.g. \cite{low2017hamiltonian, low2017optimal, low2019hamiltonian}) to implement the CTQW $U(t)=e^{iAt}$ on the graph $G=(V, E, w)$. But in order to minimize the complexity of the circuit for this task, we choose to use the simple Trotterization method which is based on the equation:
\begin{align}
e^{iAt} = \lrb{\prod_{j=1}^m e^{i w_j H_j t/N}}^N + O\lrb{\dfrac{t^2}{N}}.
\label{eq:trotter}
\end{align}
Namely, we concatenate the quantum circuits for $e^{i w_j H_j t/N}$ for $j=1,2,\dots,m$ and repeat it $N$ times. Each $e^{i w_j H_j t/N}$ can be implemented efficiently for the following reason. By conditions \ref{locality} and \ref{order} of Assumption \ref{assumption} and Eq.~\eqref{eq:def_hi}, we know that $e^{i H_j \gamma}$ either acts non-trivially on $O(\log{n})$ qubits, or is a controlled version of such an  operation (i.e. up to a permutation of the $n$ qubits, $e^{i H_j \gamma}$ is equivalent to $\ket{y}\bra{y}\otimes V \otimes I + (I - \ket{y}\bra{y}) \otimes I \otimes I$ for some $y \in \zo^q$ for some $q \in [n]$ and $O(\log{n})$-qubit unitary operation $V$.) In both cases, $e^{i H_j \gamma}$ can be implemented with $\poly{n}$ elementary gates (with the help of at most $n-1$ qubits) for arbitrary $\gamma \in \R$. It follows that the final circuit for the CTQW has $\poly{n}$ depth assuming $N=\poly{n}$.

In fact, for our purpose, it is not necessary to have faithful implementation of $U(t)=e^{iAt}$. Essentially, we want a unitary operation $W$ such that for any $j \in [k]$ and $z \in F_j$, $W \ket{z}$ is a superposition of the elements in $F_j$, and $|\bra{x} W \ket{z}|$ is large only if $x$ is close to $z$. One can see that 
\begin{align}
U'(t)\defeq \lrb{\prod_{j=1}^m e^{i w_j H_j t/N}}^N    
\end{align}
for a small $t$ satisfies this crucial property. So we can use it to replace $U(t)$ when the latter is  expensive to implement exactly.

Note, however, that a difference between $U(t)$ and $U'(t)$ is that for any $j \in [k]$ and $x, z \in F_j$, we have $\bra{x} U(t) \ket{z} \neq 0$ for generic $t$, whereas $\bra{x} U'(t) \ket{z} \neq 0$ only if the distance between $x$ and $z$ in $G$ is at most $Nm$. In other words, the ideal CTQW starting at $z$ can explore the whole connected component of $G$ that $z$ lies in, while its approximation can only explore $z$'s neighborhood in $G$ unless $Nm$ is sufficiently large. Nevertheless, Eq.~\eqref{eq:trotter} implies that when $t$ is small, $U(t)$ and $U'(t)$ do not differ much. So in this case, if $x$ is far from $z$, then $|\bra{x} U(t) \ket{z}|$ is small albeit nonzero. Namely, although the ideal CTQW starting at $z$ can explore the region far from $z$ in this case, the chance of it ending up in this region is quite small.

Both the ideal CTQW and its approximation can only explore the connected component of $G$ that the seed lies in. Therefore, if we want to fully explore the domain $F$, we need to run $U(t)$ or $U'(t)$ on $k$ different seeds, each from $F_1, F_2, \dots, F_k$ respectively. This strategy is efficient because by assumption $k=\poly{n}$. 

\subsection{Optimizing the parameters}
\label{subsec:tune_params_ctqw}
Finally, we need to tune the parameters $t$ and $\theta$ to maximize the probability of obtaining high-quality solutions from $U_{\theta}(t)\ket{z}= e^{iA_{\theta} t}\ket{z}$, where the subscript $\theta$ is added to $A$ and $U$ to emphasize their dependence on $\theta$. This is accomplished via CVaR minimization. Specifically, a measurement on $U_{\theta}(t)\ket{z}$ in the computational basis yields a random solution $X_{t, \theta}$ in $F$ such that 
\begin{align}
    \P{X_{t, \theta}=x} = \abs{\bra{x}U_{\theta}(t)\ket{z}}^2, \quad \forall x \in F.
\end{align}
Then we set $t$ and $\theta$ to be the solution of the following problem:
\begin{align}
    \min_{t, \theta \in \R} \cvar{f(X_{t, \theta})}
    \label{eq:cvar_ctqw}
\end{align}
This is done by an iterative algorithm in which each evaluation of $\cvar{f(X_{t, \theta})}$ requires to run the CTQW on a quantum device or to simulate it on a classical computer. Note that when $\alpha=1$, we have $\cvar{f(X_{t, \theta})}=\E{f(X_{t, \theta})}$. So in this case, we simply minimize the average cost of a random solution $X_{t,\theta}$ sampled from $U_{\theta}(t)\ket{z}$. The optimal choice of $\alpha$ depends on $f$ and $z$ \footnote{In our experiments, we set $\alpha=0.5$ and find that this leads to satisfactory performance}.

\section{Applications}
\label{sec:applications}
In this section, we apply CBQOA to two important combinatorial optimization problems -- Max 3SAT which is unconstrained and Max Bisection which is constrained.

\subsection{Max 3SAT}
In the MAX 3SAT problem, we are given $m$ disjunctive clauses over $n$ Boolean variables, where each clause contains at most $3$ literals, and need to find a variable assignment that maximizes the number of satisfied clauses. Here we consider the weighted version of Max 3SAT where each clause is assigned a nonnegative weight and the goal is to maximize the total weight of satisfied clauses. 

Formally, given an instance of Max 3SAT over $n$ variables $x_1, x_2, \dots, x_n$, we introduce $n+1$ auxiliary variables $x_0$, $x_{n+1}$, $x_{n+2}$, $\dots$, $x_{2n}$. A valid assignment $x=\lrb{x_0, x_1, \dots, x_{2n}}$ is a $0$-$1$ vector such that $x_0=0$ and $x_{n+i}=\neg x_i$ for $i=1,2,\dots,n$. Then every clause with at most $3$ literals can be written as $x_i \lor x_j \lor x_k$ for some $0 \le i\le j \le k \le 2n$, and it is satisfied by $x$ if and only if at least one of $x_i$, $x_j$ and $x_k$ is assigned the value $1$. We say that $(i, j, k)$ is the label of the clause $x_i \lor x_j \lor x_k$. Note that $i=0$ means that this clause has effective length $2$, $1$ or even $0$. Let $\mathcal{C}$ be the set of the labels of the clauses in a Max 3SAT instance, and for any $(i, j, k) \in \mathcal{C}$, let $w_{i,j,k}$ be the weight of the clause $x_i \lor x_j \lor x_k$ in the instance. Then we need to find a valid assignment $x=(x_0,x_1,\dots,x_{2n})$ that maximizes $\operatorname{weight}(x) \defeq \sum_{(i, j, k) \in \mathcal{C}} w_{i,j,k} \lrb{x_i \lor x_j \lor x_k}$.

\subsubsection{Generating the seed}
Building upon previous work on the PCP theorem \cite{arora1998proof}, H\r{a}stad \cite{haastad2001some} proved that assuming $P \neq NP$, no polynomial-time classical algorithm for Max 3SAT can achieve an approximation ratio exceeding $7/8$, even when restricted to satisfiable instances of the problem in which each clause contains exactly three literals. Remarkably, Karloff and Zwick \cite{karloff19977} developed a classical algorithm that achieves this ratio on satisfiable instances, which is optimal. Furthermore, there is strong evidence that their algorithm performs equally well on arbitrary Max 3SAT instances. So we use this algorithm to generate the seed for CBQOA. 

The Karloff-Zwick algorithm is based on the following SDP relaxation of the Max 3SAT instance in which each variable $x_i$ is associated with a vector $v_i$ in the unit $n$-sphere $S^n$:
\begin{align}
\quad \quad    &\max \quad \sum_{(i,j,k) \in \mathcal{C}} w_{i,j,k} z_{i, j, k},& \\
\quad \quad    &\operatorname{subject~to}  & \nonumber \\
\quad \quad    &\quad z_{i,j,k} \le \dfrac{4 - (v_0+v_i)\cdot (v_j+v_k)}{4}, & \forall (i,j,k) \in \mathcal{C},\\
\quad \quad    &\quad z_{i,j,k} \le \dfrac{4 - (v_0+v_j)\cdot (v_i+v_k)}{4}, & \forall (i,j,k) \in \mathcal{C},\\
\quad \quad    &\quad z_{i,j,k} \le \dfrac{4 - (v_0+v_k)\cdot (v_i+v_j)}{4}, & \forall (i,j,k) \in \mathcal{C},\\    
\quad \quad    &\quad z_{i,j,k} \le 1, & \forall (i,j,k) \in \mathcal{C}, \\
\quad \quad    &\quad v_i \cdot v_i = 1, & \forall 0 \le i \le 2n, \\
\quad \quad    &\quad v_i \cdot v_{n+i} = -1, & \forall 1 \le i \le n.
\end{align}
This SDP can be solved in $\tilde{O}((|\mathcal{C}|+n)^{3.5})$ time by the latest interior point method \cite{jiang2020faster}. After receiving the solution $v_0, v_1, \dots, v_{2n}$, the algorithm uses the \emph{random hyperplane rounding} technique of Goemans and Williamson \cite{goemans1995improved} to convert these vectors to a truth assignment to $x_1, x_2, \dots, x_n$. Specifically, this procedure works as follows:
\begin{enumerate}
    \item Pick a random vector $v \in S^n$ (or alternatively, chooses a random vector $v$ with $n$-dimensional standard normal distribution).
    \item For each $i \in [n]$, the variable $x_i$ is assigned value $1$ if $(v \cdot v_i)(v \cdot v_0) \ge 0$, and $0$ otherwise.
\end{enumerate} 
See Ref. \cite{karloff19977} for more details about the Karloff-Zwick algorithm.

\subsubsection{Designing and implementing the CTQW}
\label{subsubsec:sat_ctqw}
In Max 3SAT, the domain is $F=\zon$ and the cost function is $f: \zon \to \R$ such that $f(x)= \sum_{(i,j,k) \in \mathcal{C}} w_{i,j,k} \lrsb{(1-x_i)(1-x_j)(1-x_k)-1}$ (recall that $x_{n+i}=1-x_i$ for each $i \in [n]$). Suppose $z$ is a solution produced by the Karloff-Zwick algorithm. As mentioned in Section \ref{subsec:build_graph}, we define $\mathcal{T}=\lrcb{\tau_1,\tau_2,\dots,\tau_n}$, where
$\tau_i(x)=(x_1,\dots,x_{i-1}, \neg x_i, x_{i+1}, \dots, x_n)$ for all $x \in \zon$. Namely, $\tau_i$ flips the $i$-th bit of the input string. Then the term $H_i$ corresponding to $\tau_i$ is $H_i = X_i$, i.e. the Pauli X operator on the $i$-th qubit. It follows that 
\begin{align}
A_{\theta}=\sum_{i=1}^n w_i(\theta) X_i,
\end{align}
where 
\begin{align}
w_i(\theta) = \dfrac{1}{ 1 + e^{- \theta \lrsb{f(z) - f(\tau_i(z))}}}
\end{align}
depends on a parameter $\theta$. The corresponding graph $G$ is a weighted $n$-dimensional hypercube, where two vertices $x$ and $y$ are connected if and only if their Hamming distance is $1$. The CTQW on this graph is
\begin{align}
e^{iA_{\theta}t} =  \bigotimes_{j=1}^n e^{i w_j(\theta) t X} 
\label{eq:ctqw_unconstrained}
\end{align}
which can be implemented exactly by performing $e^{i w_j(\theta) t X}$ on the $j$-th qubit for $j=1,2,\dots,n$ simultaneously. The parameters $t$ and $\theta$ are optimized as mentioned in Section \ref{subsec:tune_params_ctqw}.

\subsubsection{Finishing the circuit}
In Max 3SAT, the cost function $f(x)= \sum_{(i,j,k) \in \mathcal{C}} w_{i,j,k} \lrsb{(1-x_i)(1-x_j)(1-x_k)-1}$ is a cubic polynomial in the variables $x_1,x_2,\dots,x_n$. It is mapped to the $n$-qubit Ising Hamiltonian 
\begin{align}
H_f=\sum_{(i,j,k) \in \mathcal{C}} w_{i,j,k} \lrsb{\frac{1}{8}\lrb{{I+B_i}} \lrb{{I+B_j}} \lrb{{I+B_k}} - I},    
\end{align}
where $B_i=Z_i$ if $i\le n$ and $-Z_{i-n}$ otherwise. The phase separator $e^{-iH_f\gamma}$ can be implemented with $O(|\mathcal{C}|)$ elementary gates for arbitrary $\gamma \in \R$.
 
With the methods for generating $z$ and implementing $e^{i A_\theta t}$ and $e^{-iH_f\gamma}$ in hand, we can now build the CBQOA circuit for Max 3SAT. Then we tune the parameters $(\vec \beta, \vec \gamma)$ as described in Section \ref{sec:overview}.

\subsection{Max Bisection}
In the Max Bisection problem, we are given a graph $G=(V, E, w)$ such that $|V|=n$ is even and $w: E \to \R$ assigns a weight to each edge, and need to find a subset $S \subset V$ such that $|S|=n/2$ and the total weight of the edges between $S$ and $V\setminus S$ is maximized. Namely, Max Bisection is almost the same as Max Cut, except that it has the extra constraint $|S|=|V \setminus S|$. Without loss of generality, we assume $V=[n]$ from now on.

\subsubsection{Generating the seed}
Despite the similarity between Max Cut and Max Bisection, one cannot directly use the Goemans-Williamson algorithm \cite{goemans1995improved} to solve the latter problem, because the cut found by this algorithm is not necessarily a bisection. Nevertheless, several variants of this algorithm \cite{frieze1997improved, feige2006rpr2, halperin2002unified, raghavendra2012approximating,  austrin2016better} have been proposed to address this issue. These variants typically use similar SDP relaxations but more sophisticated rounding techniques. Here we utilize one of them -- the algorithm invented by Feige and Langberg \cite{feige2006rpr2} -- to generate the seed for CBQOA. This algorithm is simple and fast, and achieves approximation ratio $0.7017$ which is satisfactory. We remark that there are other polynomial-time classical algorithms that achieve higher approximation ratios than the Feige-Langberg algorithm. In particular, the algorithm of Austrin et al. \cite{austrin2016better} achieves approximation ratio $0.8776$, which is nearly optimal under the Unique Game Conjecture \cite{khot2007optimal} (which implies that Max Bisection is NP-hard to approximate within a factor $\approx 0.8786$). However, those algorithms are slow in practice, because their time complexities are high-degree polynomials of the problem size (e.g. in order to obtain the best approximation ratio, the algorithm of Austrin et al. is estimated to take $O(n^{10^{100}})$ time). For this reason, we choose the non-optimal but more practical Feige-Langberg algorithm as the seed generator for CBQOA.

The Feige-Langberg algorithm is based on the following SDP relaxation of the Max Bisection instance in which each vertex $i \in V$ is associated with a vector $v_i$ in the unit $n$-sphere $S^n$:
\begin{align}
\quad \quad & \max \quad \dfrac{1}{2}\sum_{(i,j) \in E} w_{i,j} (1 - v_i \cdot v_j), &\\  
\quad \quad & \operatorname{subject~to} \nonumber  &\\
\quad \quad & \quad \sum_{1 \le i<j \le n} v_i \cdot v_j = -\dfrac{n}{2},  \\
\quad \quad & \quad v_i \cdot v_i = 1,   & \forall  1 \le i \le n.
\end{align}
This SDP can be solved in $\tilde{O}(n^{3.5})$ time by the latest interior point method \cite{jiang2020faster}. After obtaining the solution $v_1, v_2, \dots, v_n$, the algorithm uses a \emph{random projection, randomized rounding} ($RPR^2$) procedure \cite{feige2006rpr2} to convert these vectors to a bisection of the graph. Specifically, for arbitrary $s \in \R^+$, we define an $s$-linear function $f^*_s: \R \to [0, 1]$ as:
\begin{align}
    f^*_s(x)= \begin{cases}
    0, & \forall x \le -s; \\
    \frac{1}{2}+\frac{x}{2s}, & \forall x \in (-s, s); \\
    1, & \forall x \ge s.
    \end{cases}
\end{align}
Then the $RPR^2$ procedure works as follows:
\begin{enumerate}
    \item Let $S=\emptyset$ and $s=0.605$. 
    \item Choose a random vector $v$ with $n$-dimensional standard normal distribution.
    \item For each $i \in V$, compute $x_i = v_i \cdot v$, and add $i$ to $S$ independently with probability $f^*_s(x_i)$. 
    \item Let $S_t$ be the larger of $S$ and $V \setminus S$. For each $i \in S_t$, let $\zeta(i)=\sum_{j \in V \setminus S_t} w_{i,j}$. Suppose $S_t=\lrcb{i_1,i_2,\dots,i_l}$, where $\zeta(i_1) \ge \zeta(i_2) \ge \dots \ge \zeta(i_l)$ for some $l \ge n/2$. Then the bisection will be $\lrb{\tilde{S}_t, V \setminus \tilde{S}_t}$, where $\tilde{S}_t=\lrcb{i_1,i_2,\dots,i_{n/2}}$.
\end{enumerate}
See Ref. \cite{feige2006rpr2} for more details about the Feige-Langberg algorithm. 

\subsubsection{Designing and implementing the CTQW}
\label{subsubsec:bisection_ctqw}
\begin{figure}[ht]
    \centering
    \includegraphics[width=\linewidth]{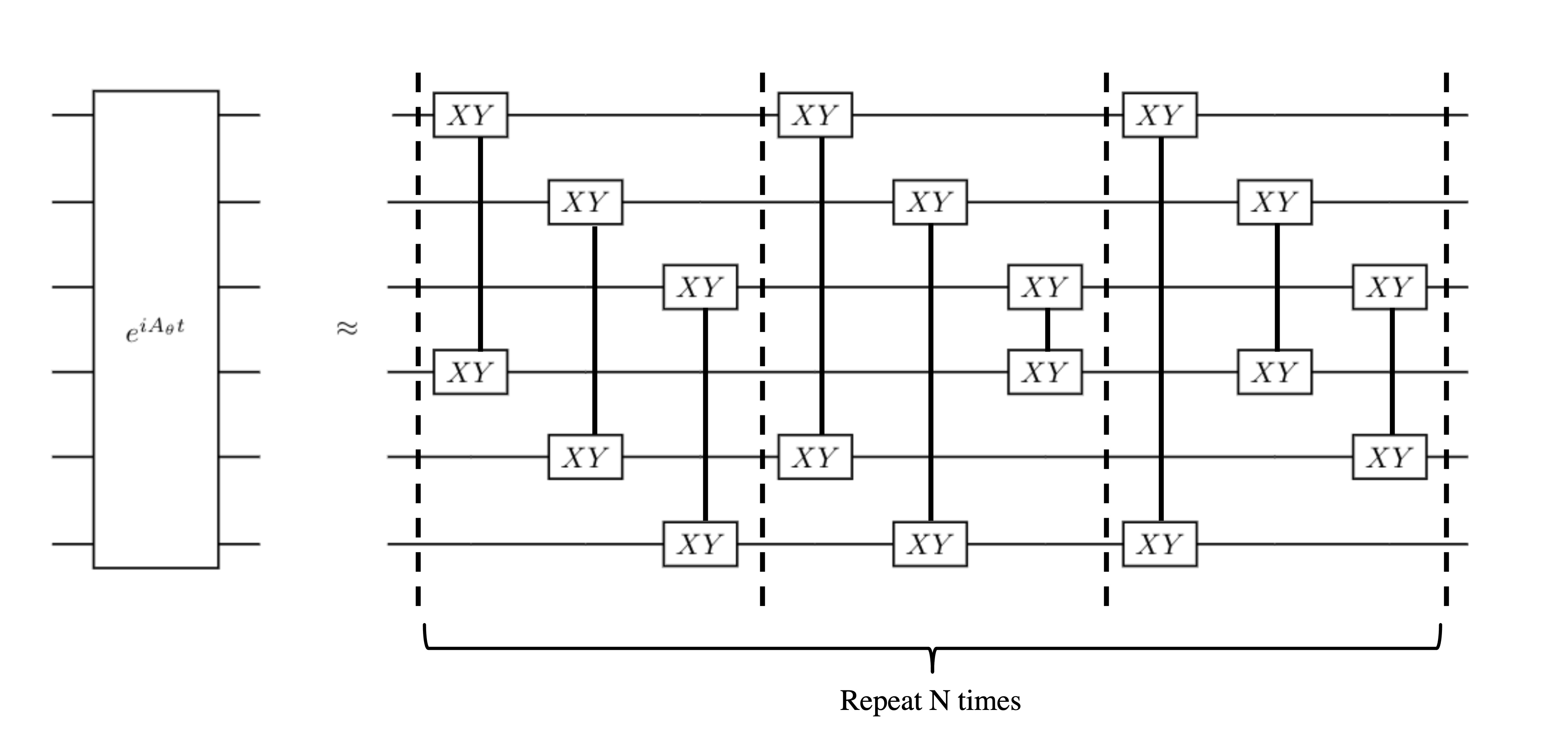}
    \caption{Quantum circuit for approximately implementing the CTQW $e^{iA_\theta t}$ for Max Bisection in the case $n=6$ and $z=000111$. Here every two $XY$ blocks connected by a vertical line represent an $XY$ gate (i.e. a unitary operation of the form $e^{i\phi(X \otimes X+Y \otimes Y)}$) on the two relevant qubits. We can perform the three $XY$ gates within the same slice simultaneously, as they act on disjoint qubits. So the depth of each layer (which is shown here) is $3$ instead of $9$. There are $N$ layers in the final circuit.}
    \label{fig:ctqw_bisection}
\end{figure}

In Max Bisection, the bisections are represented by the $n$-bit strings with Hamming weight $n/2$, and the domain is $F=\lrcb{x\in \zon: \linebreak
\sum_{i=1}^n x_i=n/2}$. The cost function is $f: F \to \R$ such that
 $f(x)=\sum_{(a,b) \in E} w_{a,b} (2 x_a x_b - x_a - x_b)$.  Since $F$ is invariant under the permutation of the $n$ bits, we can use the method in Section \ref{assumption} to construct $\mathcal{T}$. Specifically, suppose $z \in F$ is a solution produced by the Feige-Langberg algorithm. Let $T=\supp{z}$, and let $\tilde{G}=([n], \tilde{E})$ be the complete bipartite graph between $T$ and $T^c$, i.e. $\tilde{E}=\lrcb{(i, j):~i \in T,~j\in T^c}$. This leads to  $\mathcal{T}=\lrcb{\tau_{a,b}:~a\in T,~b \in T^c}$, where $\tau_{a,b}$ swaps the $a$-th and $b$-th bits of the input string. Namely, for arbitrary $x \in \zon$, we have $(\tau_{a,b}(x))_a=x_b$, $(\tau_{a,b}(x))_b=x_a$, 
and $(\tau_{a,b}(x))_c=x_c$ for all $c \in [n] \setminus \lrcb{a,b}$. Then the term $H_{a,b}$ corresponding to $\tau_{a,b}$ is 
\begin{align}
{H}_{a,b} = \sum_{x \in \zon:x_a \neq x_b} \ket{\tau_{a,b}(x)} \bra{x}
= \dfrac{1}{2} \lrb{X_a X_b + Y_a Y_b},
\end{align}
where $X_a$ and $X_b$ are the Pauli X operators on the $a$-th qubit and $b$-th qubit,  respectively, and similarly for $Y_a$ and $Y_b$. It follows that
\begin{align}
{A}_{\theta} = \sum_{a \in T}\sum_{b \in T^c} w_{a,b}(\theta) H_{a,b}
=\dfrac{1}{2}\sum_{a \in T}\sum_{b \in T^c} w_{a,b}(\theta) \lrb{X_aX_b + Y_a Y_b}.  
\end{align}
where
\begin{align}
w_{a,b}(\theta) = \dfrac{1}{ 1 + e^{- \theta \lrsb{f(z) - f(\tau_{a,b}(z))}}}
\end{align}
depends on a parameter $\theta$. Then the CTQW $e^{i{A}_{\theta}t}$ can be implemented approximately based on the equation:
\begin{align}
    e^{i{A}_{\theta}t} = \lrsb{ \prod_{a \in T} \prod_{b \in T^c} e^{i w_{a,b}(\theta) \lrb{X_aX_b + Y_a Y_b} t/{(2N)} } }^N + O\lrb{\dfrac{t^2}{N}},
    \label{eq:bisec_qw_approx}
\end{align}
where each unitary operation $e^{i w_{a,b}(\theta)  \lrb{X_aX_b + Y_a Y_b} t /{(2N)}}$ can be implemented as an $XY$ gate on the $a$-th and $b$-th qubits. The resultant circuit has $O(n^2N)$ depth. In fact, we can change the order of the terms $e^{i w_{a,b}(\theta) \lrb{X_aX_b + Y_a Y_b} t /{(2N)}}$'s to reduce the circuit depth. Specifically, suppose $T=\lrcb{a_0,a_1,\dots,a_{n/2-1}}$ and $T^c=\lrcb{b_0,b_1,\dots,b_{n/2-1}}$. Let $R_k = \lrcb{(a_i,b_{i+k \mod n/2}):~0 \le i \le n/2-1}$ for $k=0,1,\dots,n/2-1$. Then we have
\begin{align}
e^{i{A}_{\theta}t} = \lrsb{\prod_{k=0}^{n/2-1} \lrb{\prod_{(a,b)\in R_k} e^{i w_{a,b}(\theta) \lrb{X_aX_b + Y_a Y_b} t/{(2N)} }}}^N + O\lrb{\dfrac{t^2}{N}}.
    \label{eq:bisec_qw_approx2}
\end{align}
Since the unitary operations $\lrcb{e^{i w_{a,b}(\theta) \lrb{X_aX_b + Y_a Y_b} t /{(2N)}}:~(a,b)\in R_k}$ act on disjoint qubits, we can perform them simultaneously. The resultant circuit has $O(nN)$ depth. Figure \ref{fig:ctqw_bisection} illustrates this circuit for the case $n=6$ and $z=000111$. Finally, we tune the parameters $t$ and $\theta$ as described in Section \ref{subsec:tune_params_ctqw}.

\subsubsection{Finishing the circuit}
In Max Bisection, the cost function $f(x)=\sum_{(a,b) \in E} w_{a,b} (2 x_a x_b - x_a - x_b)$ is a quadratic polynomial in the variables $x_1,x_2,\dots,x_n$. It is mapped to the $n$-qubit Ising Hamiltonian
\begin{align}
    H_f = \dfrac{1}{2}\sum_{(a,b)\in E}w_{a,b}\lrb{Z_a Z_b - I}.
\end{align}
The phase separator $e^{-iH_f \gamma}$ can be implemented with $|E|$ $ZZ$ gates (i.e. unitary operations of the form $e^{i\phi Z \otimes Z}$) for arbitrary $\gamma \in \R$.

Equipped with the means to generate $z$ and implement $e^{i A_\theta t}$ and $e^{-iH_f\gamma}$, we can  now build the CBQOA circuit for Max Bisection. Then we optimize the parameters $(\vec \beta, \vec \gamma)$ as mentioned in Section \ref{sec:overview}.

\section{Experimental evaluation}
\label{sec:experiments}
In this section, we assess the performance of CBQOA on Max 3SAT and Max Bisection. We utilize the Karloff-Zwick and Feige-Langberg algorithms to generate the seeds for CBQOA in these problems, respectively, and find that CBQOA yields better solutions than the seeds it receives with high probability. Moreover, the simulation results also indicate that CBQOA outperforms GM-QAOA on these problems. 

\subsection{Performance metrics}
Traditionally, the approximation ratio of a solution to an optimization problem is defined as the ratio of the objective value of this solution to that of the optimum solution. Namely, suppose we want to maximize or minimize the objective function $f: F \subseteq \zon \to \mathbb{R}$. The approximation ratio of a solution $z \in F$ is often defined as
\begin{align}
\alpha(z)\defeq \dfrac{f(z)}{f(z^*)},
\label{eq:approx_ratio1}
\end{align}
where $z^* \in F$ is an optimum solution. However, if $f(z)$ can be positive or negative or zero, then this definition is problematic and needs modification. So here we introduce an alternative metric for measuring the quality of a solution $z \in F$: 
\begin{align}
\beta(z)\defeq\dfrac{\mathbb{E}_x[f(x)]-f(z)}{\mathbb{E}_x[f(x)]-f(z^*)},
\label{eq:approx_ratio2}
\end{align}
where $x$ is chosen uniformly at random from $F$. This metric is usable in all scenarios. One can see that $\beta(z)$ measures to what extent $z$ is better than a random guess. Note that $\beta(z) \le 1$ for all $z \in F$, and the equality holds if and only if $f(z)=f(z^*)$. Moreover, $\beta(z)$ can be negative, which happens if and only if $z$ has lower quality than the average quality of a random guess. Finally, $\beta(z)$ remains intact if we replace $f$ by $c f + d$ for arbitrary $c \in \R^+$ and $d \in \R$. This is consistent with the intuition that the quality of a solution $z \in F$ should remain invariant under such transformations of the objective function. \emph{From now on, we will call $\beta(z)$ the approximation ratio of a  solution $z \in F$}.

Given a classical or quantum algorithm $\mathcal{A}$ for solving an optimization problem, we measure its performance by the probability of it producing ``good" solutions. Here we say that a solution $z \in F$ is \emph{good} if its approximation ratio $\beta(z)$ exceeds certain threshold. Formally, let $\pogs_x(\mathcal{A})$ be the probability of $\mathcal{A}$ producing a solution $z \in F$ such that $\beta(z) \ge x$, where $x \in (-\infty, 1]$ is the threshold for the approximation ratio. Here $\pogs$ is short for \emph{probability of good solutions}. Furthermore, we could repeat $\mathcal{A}$ multiple times and reduce the failure probability (i.e. the probability of not receiving a good solution) exponentially. Namely, let $\mathcal{A}^k$ denote the event that we run $\mathcal{A}$ $k$ times independently, obtain $k$ solutions, and pick the best one of them. Then we have 
\begin{align}
    \pogs_x(\mathcal{A}^k) = 1 - (1-\pogs_x(\mathcal{A}))^k.
\end{align} 

\subsection{Experimental setup}
For convenience, we will use KZ and FL to denote the Karloff-Zwick and Feige-Lanberg algorithms, respectively, and use $\gmqaoa_p$ and $\cbqoa_p$ to denote the $p$-layer GM-QAOA and $p$-layer CBQOA algorithms, respectively. 

We will focus on the hard instances of Max 3SAT and Max Bisection on which KZ and FL do not perform well and check to what extent CBQOA enhances the outputs of these classical algorithms. Moreover, we also compare the performance of GM-QAOA and CBQOA on these instances.

Specifically, we generate $100$ hard Max 3SAT instances as follows. First, we create a random Max 3SAT instance with $16$ variables and $200$ clauses, and assign a uniformly random weight between $0$ and $1$ to each clause independently. Then we solve the Karloff-Zwick SDP for this instance, and perform the random hyperplane rounding on the solution $10000$ times, receiving $10000$ variable assignments. Then the fraction of the variable assignments with  approximation ratios at least $0.7$ is an accurate estimate of $\pogs_{0.7}(\kz)$ with high probability. If this quantity is smaller than $0.05$, then this Max 3SAT instance is deemed hard and added to our benchmark set. We repeat this process until $100$ such instances are collected. 

Similarly, we generate $100$ hard Max Bisection instances as follows. First, we create an Erd\H{o}s-R\'{e}nyi random graph $G(n, p)$ with $n=12$ vertices and edge probability $p=0.5$, and assign a uniformly random weight between $-1$ and $1$ to each edge independently. Then we solve the Feige-Langberg SDP for this instance, and perform the $RPR^2$ procedure on the solution $10000$ times, receiving $10000$ bisections. Then the fraction of the bisections with approximation ratios at least $0.99$ is an accurate estimate of $\pogs_{0.99}(\fl)$ with high probability \footnote{We find that FL performs extremely well on small graphs. So we have to focus on the probabilities of the classical and quantum algorithms producing nearly-optimal solutions in this case. We expect that as the graph size grows, the performance of FL will deteriorate, and the gaps between the $\pogs_x$ of FL and CBQOA will widen even for $x$ far from $1$. However, it is quite time-consuming to simulate the quantum circuits for such graphs on classical computers, and we leave it as future work to verify this conjecture.}. If this quantity is smaller than $0.05$, then this Max Bisection instance is deemed hard and added to our benchmark set. We repeat this process until $100$ such instances are gathered. 

In CBQOA, we use KZ and FL to generate the seeds for Max 3SAT and Max Bisection, respectively. We implement the CTQW for Max 3SAT exactly as mentioned in Section \ref{subsubsec:sat_ctqw}, and implement the CTQW for Max Bisection approximately as described in Section \ref{subsubsec:bisection_ctqw}, setting $N=3$. Then we tune the parameters $\theta$ and $t$ as mentioned in Section \ref{subsec:tune_params_ctqw}, setting $\alpha=0.5$. Finally, we optimize the parameters $(\vec \beta, \vec \gamma)$ as described in Section \ref{sec:overview}, again setting $\alpha=0.5$. During this procedure, we utilize the technique in Appendix \ref{subsec:simulate_cbqoa_circuit} to accelerate the classical simulation of CBQOA circuits, setting $M=1000$ (which leads to sufficiently accurate results). All the CVaR minimization problems were solved by the ADAM optimizer. All the simulations were conducted in the Orquestra$^\text{\textregistered}$ platform \footnote{\href{https://www.orquestra.io/}{https://www.orquestra.io/}}.

\subsection{Simulation results}
\begin{figure}[htbp]
    \centering
    \includegraphics[width=0.8\textwidth]{"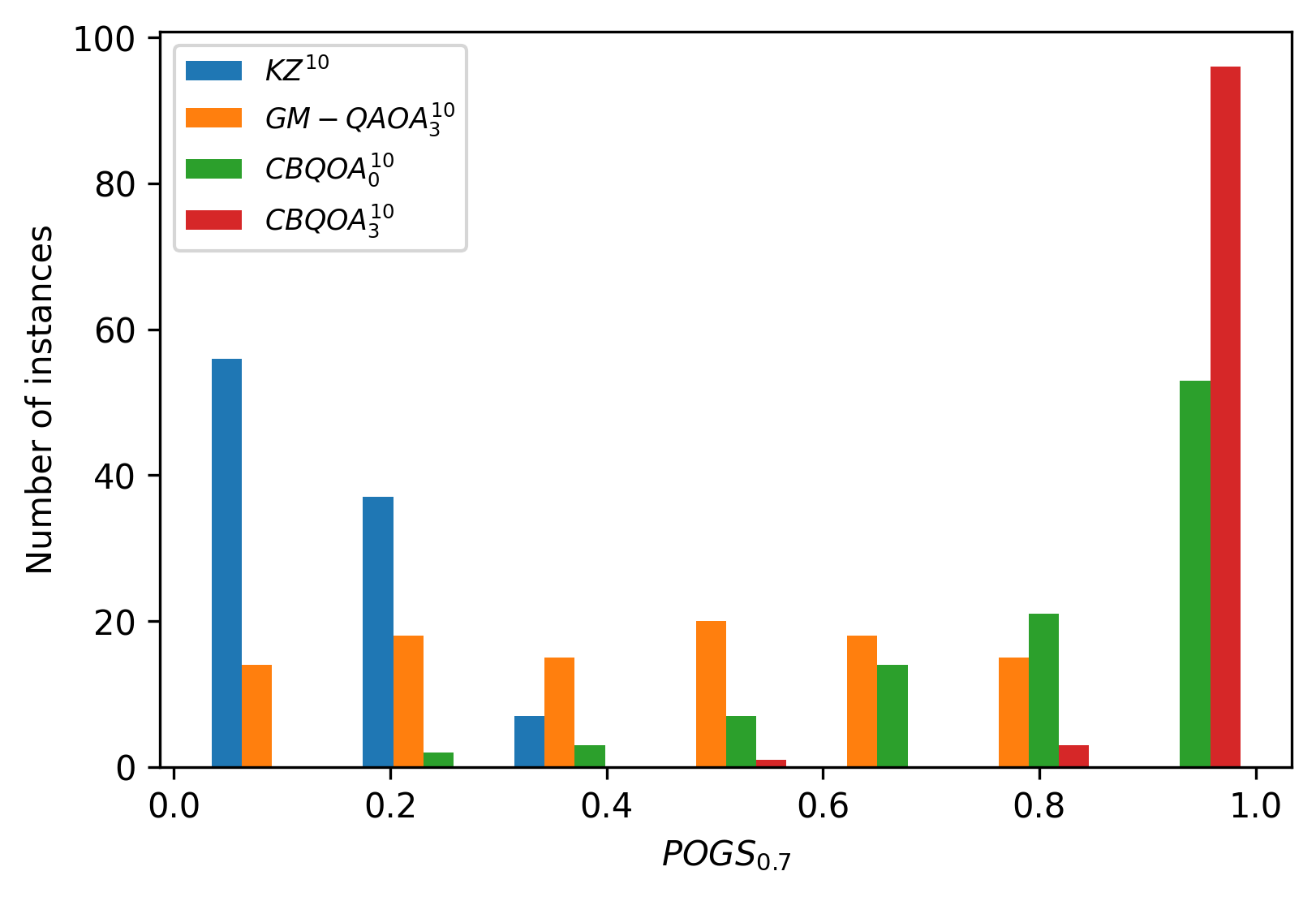"}
    \caption{This figure illustrates the histograms of the $\pogs_{0.7}$ of $\fl^5$, $\gmqaoa_3^5$, $\cbqoa_0^5$ and $\cbqoa_3^5$ on the $100$ hard Max 3SAT instances.}
    \label{fig:max_3sat_0.7}
\end{figure}

\begin{figure}[htbp]
    \centering
    \includegraphics[width=0.8\textwidth]{"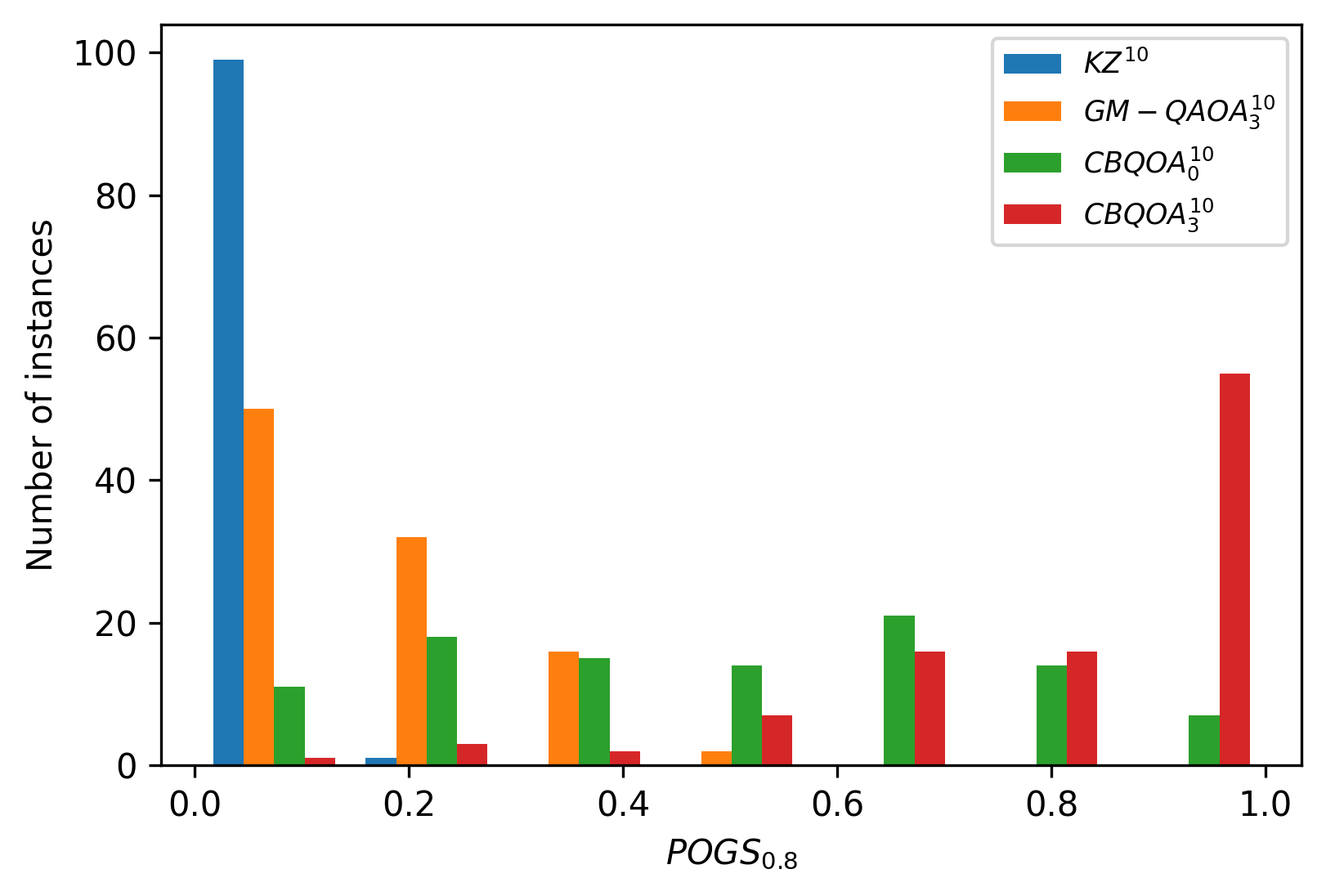"}
    \caption{This figure illustrates the histograms of the $\pogs_{0.8}$ of $\fl^5$, $\gmqaoa_3^5$, $\cbqoa_0^5$ and $\cbqoa_3^5$ on the $100$ hard Max 3SAT instances.}
    \label{fig:max_3sat_0.8}
\end{figure}

Figures \ref{fig:max_3sat_0.7} and \ref{fig:max_3sat_0.8} illustrate the histograms of the $\pogs_{0.7}$ and $\pogs_{0.8}$ of $\kz^{10}$, $\gmqaoa_3^{10}$, $\cbqoa_0^{10}$ and $\cbqoa_3^{10}$ on the $100$ hard Max 3SAT instances, respectively. One can see that CBQOA$_3$ performs the best among them. For example, Figure \ref{fig:max_3sat_0.8} shows that KZ rarely produces a solution with approximation ratio at least $0.8$. However, its mediocre output and a properly-constructed CTQW enable $\cbqoa_0$ to find such a solution with noticeable probability. Then this probability is significantly amplified in $\cbqoa_3$ thanks to the 3 layers of phase separators and mixing operators in the circuit.

Figures \ref{fig:max_3sat_0.7} and \ref{fig:max_3sat_0.8} also reveal that $\gmqaoa_3$ does not perform as well as $\cbqoa_3$. In fact, $\gmqaoa_3$ is even surpassed by $\cbqoa_0$. Recall that GM-QAOA always starts with a uniform superposition of the feasible solutions (many of which have low qualities), while CBQOA starts with a non-uniform superposition of the feasible solutions in which the neighbors of the seed receive higher amplitudes than the others. As a consequence, the initial state of CBQOA has larger overlap with the states corresponding to the high-quality solutions than that of GM-QAOA, which means that CBQOA needs fewer layers than GM-QAOA to reach the same performance, as supported by the experimental data.

\begin{figure}[htbp]
    \centering
    \includegraphics[width=0.8\textwidth]{"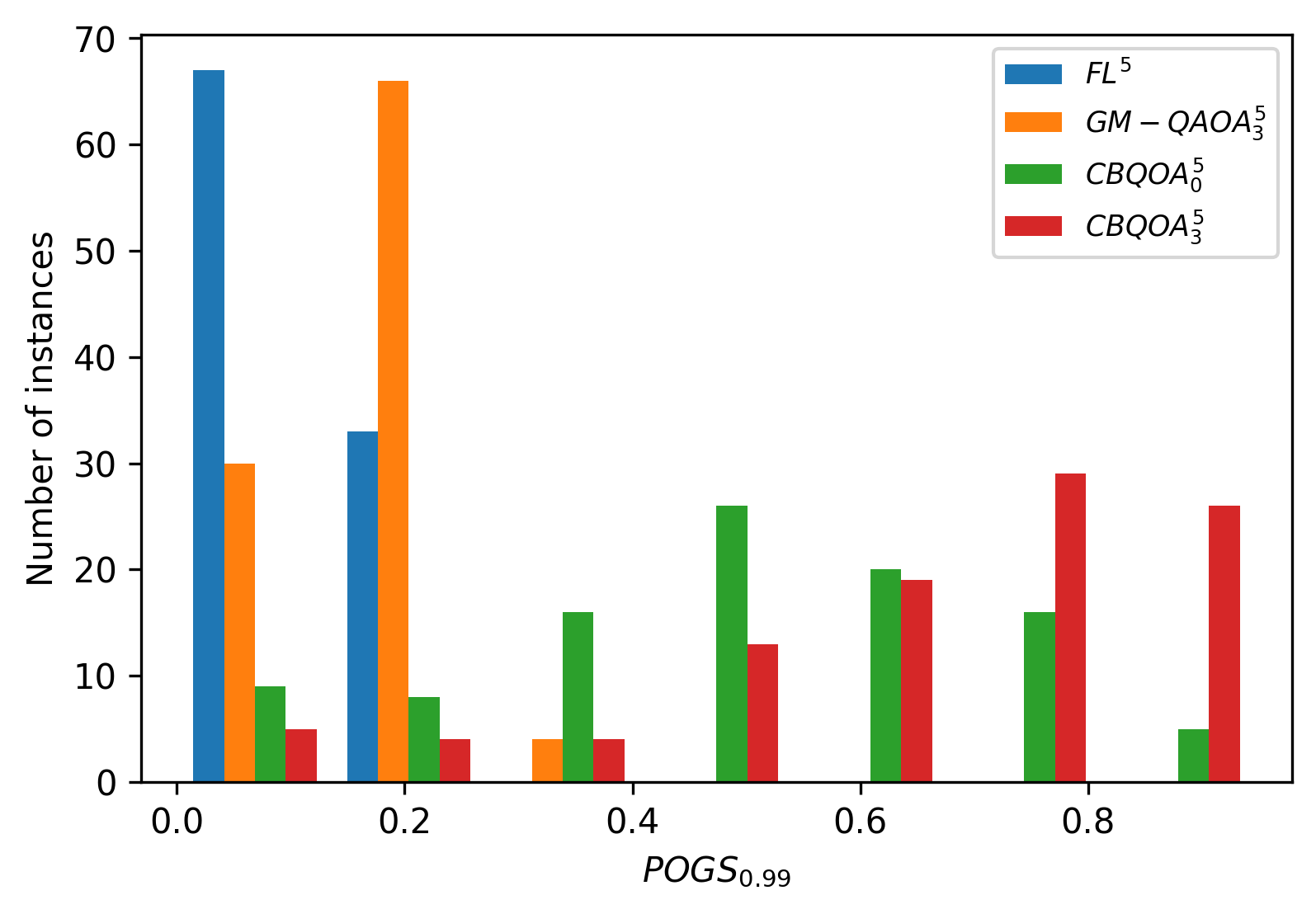"}
    \caption{This figure illustrates the histograms of the $\pogs_{0.99}$ of $\fl^5$, $\gmqaoa_3^5$, $\cbqoa_0^5$ and $\cbqoa_3^5$ on the $100$ hard Max Bisection instances. }
    \label{fig:max_bisection_0.99}
\end{figure}

Figure \ref{fig:max_bisection_0.99} illustrates the histograms of the $\pogs_{0.99}$ of $\fl^5$, $\gmqaoa_3^5$, $\cbqoa_0^5$ and $\cbqoa_3^5$ on the $100$ hard Max Bisections instances. One can see that $\cbqoa_3$ performs the best among them. Specifically, FL struggles to produce a solution with approximation ratio at least $0.99$. But with the help of its output and an appropriate CTQW, $\cbqoa_0$ finds such a solution with decent chance. Then $\cbqoa_3$ uses a 3-layer amplitude-amplification-like circuit to greatly boost this probability. Meanwhile, $\gmqaoa_3$ is again surpassed by $\cbqoa_3$, for a reason similar to the one above for Max 3SAT.

\begin{figure}[htbp]
    \centering
    \includegraphics[width=0.8\textwidth]{"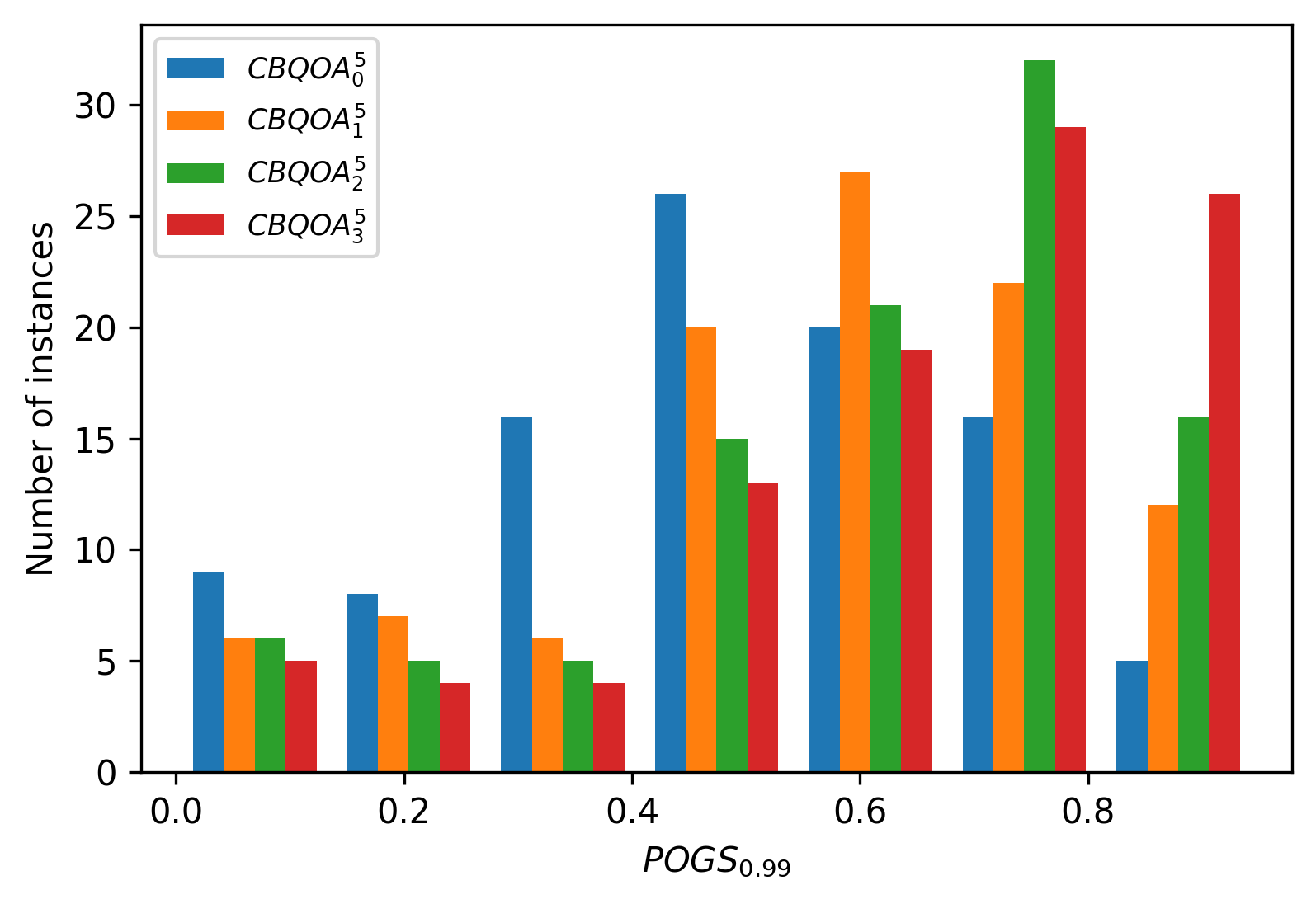"}
    \caption{This figure illustrates the histograms of the $\pogs_{0.99}$ of $\cbqoa_p^5$ for $p=0,1,2,3$ on the $100$ hard Max Bisection instances. }
    \label{fig:max_bisection_0.99_depth_impact}
\end{figure}

Figure \ref{fig:max_bisection_0.99_depth_impact} illustrates the histograms of the $\pogs_{0.99}$ of $\cbqoa_p^5$ for $p=0,1,2,3$ on the $100$ hard Max Bisections instances. One can see that deeper circuits yield better results. However, as $p$ grows large, increasing it further leads to diminishing returns while making the parameter tuning more difficult and time-consuming. So we need to choose it appropriately to achieve a balance between the efficiency of the algorithm and the quality of its output.

\section{Conclusions}
\label{sec:conclusion}
To summarize, we have proposed CBQOA -- a hybrid quantum-classical algorithm for solving a broad class of combinatorial optimization problems. This algorithm utilizes an approximate solution produced by a classical algorithm and a CTQW on a properly-constructed graph to create a suitable superposition of the feasible solutions which is then processed in certain way. This algorithm has the merits that it solves constrained problems without modifying their cost functions, confines the evolution of the quantum state to the feasible subspace, and does not rely on efficient indexing of the feasible solutions. We have demonstrated the applications of CBQOA to Max 3SAT and Max Bisection, and provided empirical evidence that CBQOA outperforms previous approaches on these problems. 

The fact that quantum computing offers novel approaches to combinatorial optimization does not mean that we should abandon existing classical techniques for the same problems altogether. For one thing, one can exploit such techniques to generate valuable information that greatly benefits quantum optimization algorithms, as shown in this work. For another, classical gates take less times to execute than quantum gates, which means that one can perform more classical operations than quantum operations within the same amount of time. So it is conceivable that the most cost-effective algorithms for tackling combinatorial optimization problems contains both classical and quantum components and delegates appropriate subtasks to them in order to achieve the highest overall performance. 

Finally, we would like to point out a few research directions that are worth further exploration:
\begin{itemize}
    \item Currently, we use a well-known classical approximation algorithm to generate the seed for CBQOA. Although this strategy proves effective, it might be not optimal. Ideally, we want to find a solution $z$ that is close to a large number of high-quality solutions, so that the state $\ket{\psi}=e^{iAt}\ket{z}$ has large overlap with the states corresponding to those solutions for some appropriate $t$. It is the distances between $z$ and the high-quality solutions, not the value of $f(z)$, that plays a significant role in CBQOA. It would be beneficial to develop a better classical algorithm for this task and utilize it in CBQOA. 
    
    \item For CBQOA to be practical, we need to make the parameter tuning process as efficient as possible. So far, we have used a general-purpose iterative optimizer to tune the parameters (i.e. $t$, $\theta$, $\vec \beta$ and $\vec \gamma$), in which each iteration requires to run the CBQOA circuit on a quantum device. It is worth investigating whether one can exploit the structure of the CBQOA circuit to reduce the number of iterations in this process, or to develop more efficient methods for this task. In particular, is it possible to design a class of parameter schedules \cite{zhou2020quantum, kremenetski2021quantum} for CBQOA that lead to excellent performance? 
    
    \item In this work, we have only considered a natural way to leverage well-established classical techniques to enhance quantum optimization. That is, we use a classical algorithm to find a feasible solution such that its neighborhood is likely to contain high-quality feasible solutions, and use a quantum circuit to search this region for such solutions. A handful of other approaches have been studied in Refs. \cite{egger2021warm, tate2020bridging, tate2021classically, van2021quantum}. We hope that these works will stimulate more effort in combining the strengths of the two computing paradigms to advance the state of the art for combinatorial optimization. 
\end{itemize}

\section*{Acknowledgement}
We thank Peter Johnson, Alejandro Perdomo-Ortiz and Sukin Sim for insightful discussions and valuable feedback, and thank Micha\l{} St\k{e}ch\l{}y, Mesut Celik and Alison Roeth for their help with running the experiments in the Orquestra$^\text{\textregistered}$ platform. 
\appendix

\section{Accelerating the classical simulation of CBQOA circuits}
\label{subsec:simulate_cbqoa_circuit}
In this appendix, we describe a technique for accelerating the simulation of CBQOA circuits on classical computers, assuming $\ket{\psi}=e^{iAt} \ket{z}$ is already known. Here our goal is to approximate the probability distribution of $f(X_{\vec \beta, \vec \gamma})$ where $X_{\vec \beta, \vec \gamma}$ is the random variable given by Eq.~\eqref{eq:prob_dist_x}. Then we can use this information to compute $\cvar{f(X_{\vec \beta,\vec \gamma})}$ and update the parameters $(\vec \beta,\vec \gamma)$ accordingly in the iterative optimizer. This technique generalizes the one for simulating Grover-Mixer Threshold QAOA (GM-Th-QAOA) circuits in Ref. \cite{golden2021threshold}.

Our basic idea is to replace $f$ with a function $\tilde{f} \approx f$ such that the CBQOA circuit for $\tilde{f}$ can be more easily simulated, and its output state is close to the one for $f$. Specifically, let $a, b \in \R$ be such that $a \le f(x) < b$ for all $x \in \zon$. Let $M \in \Z^+$ be determined later, and let $\Delta=(b-a)/M$. Let 
\begin{align}
S_j = \lrcb{ x \in \zon:~a + j \Delta \le f(x) < a + (j+1) \Delta},  
\end{align} 
for $j=0, 1,\dots, M-1$. Then $S_0, S_1, \dots, S_{M-1}$ form a partition of $\zon$. Define a function $\tilde{f}: \zon \to \R$ as follows. For any $x \in \zon$, 
\begin{align}
    \tilde{f}(x) = a + \Delta \lrb{\left \lfloor \dfrac{f(x) - a }{\Delta} \right\rfloor + \dfrac{1}{2}}.
\end{align}
In other words, $\tilde{f}(x)=\tilde{f}_j:=a + (j+1/2) \Delta$ for all $x \in S_j$, for $j=0,1,\dots,M-1$. One can see that $|f(x) - \tilde{f}(x)| \le \Delta/2$ for all $x \in 
\zon$. So $f$ and $\tilde{f}$ are close if $M$ is sufficiently large.

We encode $\tilde{f}$ into an $n$-qubit Ising Hamiltonian 
\begin{align}
H_{\tilde{f}} = \sum_{x \in \zon} \tilde{f}(x) \ketbra{x}{x}
= \sum_{j=0}^{N-1} \tilde{f}_j \lrb{\sum_{x \in S_j} \ketbra{x}{x}}.    
\label{eq:def_h_tf}
\end{align}
To avoid confusion, we will use $\ket{\psi_f(\vec \beta, \vec \gamma)}$
and $\ket{\psi_{\tilde{f}}(\vec \beta, \vec \gamma)}$ to denote the ansatz states for $f$ and $\tilde{f}$, respectively, i.e.
\begin{align}
\ket{\psi_f(\vec \beta, \vec \gamma)} = e^{-i \beta_p \ketbra{\psi}{\psi}} e^{-i \gamma_p H_f}
e^{-i \beta_{p-1} \ketbra{\psi}{\psi}} e^{-i \gamma_{p-1} H_f}\dots 
e^{-i \beta_1 \ketbra{\psi}{\psi}} e^{-i \gamma_1 H_f} \ket{\psi}, \\
\ket{\psi_{\tilde{f}}(\vec \beta, \vec \gamma)} = e^{-i \beta_p \ketbra{\psi}{\psi}} e^{-i \gamma_p H_{\tilde{f}}}
e^{-i \beta_{p-1} \ketbra{\psi}{\psi}} e^{-i \gamma_{p-1} H_{\tilde{f}}}\dots 
e^{-i \beta_1 \ketbra{\psi}{\psi}} e^{-i \gamma_1 H_{\tilde{f}}} \ket{\psi}.
\end{align}
Let $X \defeq X_{\vec \beta, \vec \gamma}$, and let $\tilde{X} \defeq \tilde{X}_{\vec \beta, \vec \gamma}$ be a random variable such that 
\begin{align}
\P{\tilde{X} = x}=|\braket{x}{\psi_{\tilde{f}}(\vec \beta, \vec \gamma)}|^2,~\forall x \in F.    
\end{align}
Then we can get the probability distribution of $\tilde{f}(\tilde{X})= \bra{\tilde{X}} H_{\tilde{f}} \ket{\tilde{X}}$ as follows. Suppose $\ket{\psi}=\sum_{x \in \zon} \alpha_x \ket{x}$ for some known $\alpha_x$'s. We can re-write it as $\ket{\psi} = \sum_{j=0}^{M-1} \eta_j \ket{\psi_j}$, where $\eta_j = \sqrt{\sum_{x \in S_j} |\alpha_x|^2}$, and $\ket{\psi_j}=\frac{1}{\eta_j}\sum_{x \in S_j} \alpha_x \ket{x}$ is normalized, for $j=0,1,\dots,M-1$. Meanwhile, for $t=0,1,\dots,p$, let 
\begin{align}
\ket{\phi^{(t)}} = e^{-i \beta_t \ketbra{\psi}{\psi}} e^{-i \gamma_t H_{\tilde{f}}}
e^{-i \beta_{p-1} \ketbra{\psi}{\psi}} e^{-i \gamma_{p-1} H_{\tilde{f}}}\dots 
e^{-i \beta_1 \ketbra{\psi}{\psi}} e^{-i \gamma_1 H_{\tilde{f}}} \ket{\psi},   
\end{align}
Then one can prove by induction that $\ket{\phi^{(t)}} \in \myspan{\ket{\psi_j}:~j=0,1,\dots,M-1}$ for all $t$'s. Furthermore, $\braket{\psi_j}{\phi^{(t)}}$ can be computed as follows. Suppose 
\begin{align}
\ket{\phi^{(t)}} = \sum_{j=0}^{M-1} \eta_j^{(t)} \ket{\psi_j}
\end{align}
for some $\eta_j^{(t)}$'s. The by Eq.~\eqref{eq:def_h_tf} and the fact that
\begin{align}
\ket{\phi^{(t)}}= e^{-i \beta_t \ketbra{\psi}{\psi}} e^{-i \gamma_t H_{\tilde{f}}}\ket{\phi^{(t-1)}},
\end{align}
one can show the following relationship between the $\eta_j^{(t)}$'s and $\eta_j^{(t-1)}$'s:
\begin{align}
\eta_j^{(t)} = \eta_j^{(t-1)} e^{-i\tilde{f}_j\gamma_t}
 + (e^{-i \beta_t}-1) \eta_j \lrb{\sum_{k=0}^{M-1} \eta_k \eta_k^{(t-1)} e^{-i \tilde{f}_k \gamma_t}},~~~\forall 0 \le j \le M-1.
 \label{eq:recursion}
\end{align}
As a consequence, we can start with $\eta^{(0)}_j = \eta_j$ for each $j$, and use Eq.~\eqref{eq:recursion} to iteratively compute the $\eta^{(1)}_j$'s, $\eta^{(2)}_j$'s, $\dots$, $\eta^{(p)}_j$'s. Finally, noting that $\ket{\phi^{(p)}}=\ket{\psi_{\tilde{f}}(\vec \beta, \vec \gamma)}$, we obtain
\begin{align}
\P{\tilde{f}(\tilde{X})=\tilde{f}_j} = |\eta^{(p)}_j|^2,~~~\forall 0 \le j \le M-1,
\label{eq:approx_prob}
\end{align}
which is the desired probability distribution. Apart from the computation of $\eta_j$'s from  $\ket{\psi}$, this procedure takes only $O(M^2p)$ time and $O(M)$ space, in contrast to the $\Omega(2^n p)$ time complexity and $\Omega(2^n)$ space complexity of the naive method.

Once we obtain the probability distribution of $f(\tilde{X})$ as in Eq.~\eqref{eq:approx_prob}, we use it to compute $\cvar{f(\tilde{X})}$ for an appropriate confidence level $\alpha$. As shown below, by choosing some $M=\poly{p, b-a, 1/\alpha, 1/\epsilon}$, we can ensure 
\begin{align}
\abs{\cvar{f(\tilde{X})} - \cvar{f(X)}} \le \epsilon.    
\end{align}
Namely, $\cvar{f(\tilde{X})}$ is an accurate estimate of $\cvar{f(X)}$, and hence we can use it in the latter's place to update the parameters $(\vec \beta, \vec \gamma)$ in the iterative optimizer. 

In many problems, we have $b-a=\poly{n}$. Moreover, we often set $p=\poly{n}$ and $\alpha=\Omega(1)$ in practice. In this case, it is sufficient to choose some $M=\Theta(\poly{n}/\epsilon)$ to guarantee that $\cvar{f(\tilde{X})}$ is $\epsilon$-close to $\cvar{f(X)}$. Consequently, apart from the computation of the $\eta_j$'s from $\ket{\psi}$, our method for simulating the CBQOA circuit takes $\Theta(\poly{n}/\epsilon)$ space and time, which is much more efficient than the naive method which takes $\Omega(2^n)$ space and time. 

\subsection{Error analysis}
\label{subsec:error_analysis}
Now we prove that if $\abs{\gamma_j}=O(1)$ for $j=1,2,\dots,p$, then there exists some $M=\poly{p, b-a, 1/\alpha, 1/\epsilon}$ such that 
\begin{align}
    \abs{\cvar{f(\tilde{X})} - \cvar{f(X)}} \le \epsilon.
\end{align}

For convenience, let $U(\beta)=e^{-i\beta \ketbra{\psi}{\psi}}$, $V(\gamma)=e^{-i \gamma H_f}$ and $\tilde{V}(\gamma)=e^{-i \gamma H_{\tilde{f}}}$ for arbitrary $\beta, \gamma \in \R$. Then we have
\begin{align}
\ket{\psi_f(\vec \beta, \vec \gamma)}= U(\beta_p)V(\gamma_p) U(\beta_{p-1})V(\gamma_{p-1})\dots U(\beta_1) V(\gamma_1) \ket{\psi},\\
\ket{\psi_{\tilde{f}}(\vec \beta, \vec \gamma)}= U(\beta_p)\tilde{V}(\gamma_p) U(\beta_{p-1})\tilde{V}(\gamma_{p-1})\dots U(\beta_1) \tilde{V}(\gamma_1) \ket{\psi}.
\end{align}
By a hybrid argument, one can show that
\begin{align}
&\norm{U(\beta_p)V(\gamma_p) U(\beta_{p-1})V(\gamma_{p-1})\dots U(\beta_1) V(\gamma_1)  - U(\beta_p)\tilde{V}(\gamma_p) U(\beta_{p-1})\tilde{V}(\gamma_{p-1})
\dots U(\beta_1) \tilde{V}(\gamma_1)  }  \\
&\le \sum_{t=1}^p \norm{V(\gamma_t) - \tilde{V}(\gamma_t)}.
\end{align}
Meanwhile, since $|\gamma_t|=O(1)$, we have
\begin{align}
\norm{V(\gamma_t) - \tilde{V}(\gamma_t)}
&=\max_{x \in \zon} |e^{-i f(x) \gamma_t} - e^{-i \tilde{f}(x) \gamma_t}|    \\
&=O(\max_{x \in \zon} |f(x) - \tilde{f}(x)|)  \\
&=O(\Delta)
\end{align}
Combining these two facts yields
\begin{align}
&\norm{U(\beta_p)V(\gamma_p) U(\beta_{p-1}){V}(\gamma_{p-1})\dots U(\beta_1) V(\gamma_1)  - 
U(\beta_p)\tilde{V}(\gamma_p)U(\beta_{p-1})\tilde{V}(\gamma_{p-1}) \dots U(\beta_1) \tilde{V}(\gamma_1) } \\
&=O(p\Delta).    
\end{align}
Then since $\norm{\ket{\psi}}=1$, we get
\begin{align}
&\norm{\ket{\psi_f(\vec \beta, \vec \gamma)} - \ket{\psi_{\tilde{f}}(\vec \beta, \vec \gamma)}} \\
&=\norm{\lrb{U(\beta_p)V(\gamma_p) U(\beta_{p-1}){V}(\gamma_{p-1}) \dots U(\beta_1) V(\gamma_1)  - 
U(\beta_p)\tilde{V}(\gamma_p) U(\beta_{p-1})\tilde{V}(\gamma_{p-1}) \dots U(\beta_1) \tilde{V}(\gamma_1)} \ket{\psi}}\\
&=O(p\Delta).
\label{eq:state_dist}
\end{align}
Let $\rho=\ketbra{\psi_f(\vec \beta, \vec \gamma)}{\psi_f(\vec \beta, \vec \gamma)}$ and 
$\sigma=\ketbra{\psi_{\tilde f}(\vec \beta, \vec \gamma)}{\psi_{\tilde f}(\vec \beta, \vec \gamma)}$. Then it follows from Eq.~\eqref{eq:state_dist} that
\begin{align}
D(\rho, \sigma)
\defeq \dfrac{1}{2} \mytr{\abs{\rho - \sigma}}
= O(p \Delta).
\label{eq:trace_dist_rho_sigma}
\end{align}

Now suppose $\zon = \lrcb{x^{(1)}, x^{(2)}, \dots, x^{(N)}}$ such that 
$f(x^{(1)}) \le f(x^{(2)}) \le \dots \le f(x^{(N)})$, where $N=2^n$.
Let $\vec p = (p_1, p_2, \dots, p_N)$ where
$p_i = \bra{x^{(i)}} \rho  \ket{x^{(i)}}$, 
and $\vec q = (q_1, q_2, \dots, q_N)$ where
$q_i = \bra{x^{(i)}} \sigma \ket{x^{(i)}}$, for $i=1,2,\dots,N$. 
Then by Eq.~\eqref{eq:trace_dist_rho_sigma} and Theorem 9.1 of Ref. \cite{nielsen2002quantum}, we get
\begin{align}
D(\vec p, \vec q) \defeq \dfrac{1}{2} \sum_{i=1}^N \abs{p_i - q_i}
\le D(\rho, \sigma) = O(p \Delta).
\label{eq:prob_dist}
\end{align}

Let $g: \zon \to \R$ be defined as $g(x)=f(x)-(a+b)/2$. Then we have $\abs{g(x)} \le (b-a)/2$ for all $x \in \zon$, and
\begin{align}
\cvar{f(X)}=\cvar{g(X)} + (a+b)/2,    \\
\cvar{f(\tilde{X})}=\cvar{g(\tilde{X})} + (a+b)/2,
\end{align}
and hence 
\begin{align}
    \cvar{f(X)} - \cvar{f(\tilde{X})} = \cvar{g(X)} - \cvar{g(\tilde{X})}.
\end{align}

Let $j$ be the unique integer in $[N]$ such that
$\sum_{i=1}^{j-1} p_i < \alpha$ and $\sum_{i=1}^j p_i \ge \alpha$. Then 
\begin{align}
    \cvar{g(X)} = \lrb{\sum_{i=1}^{j-1} p_i g(x^{(i)}) + p_j' g(x^{(j)})}/\alpha
    \label{eq:cvargx}
\end{align}
where 
$p_j'=\alpha - \sum_{i=1}^{j-1} p_i$. Similarly, let $k$ be the unique integer in $[N]$ such that
$\sum_{i=1}^{k-1} q_i < \alpha$ and $\sum_{i=1}^k q_i \ge \alpha$. Then 
\begin{align}
    \cvar{g(\tilde{X})} = \lrb{\sum_{i=1}^{k-1} q_i g(x^{(i)}) + q_k' g(x^{(k)})}/\alpha
    \label{eq:cvargtx}
\end{align}
where 
$q_k'=\alpha - \sum_{i=1}^{k-1} q_i$. In order to bound $\abs{\cvar{g(\tilde{X})} - \cvar{g(X)}}$, we consider the cases $j<k$, $j=k$ and $j>k$ separately.
\begin{itemize}
    \item Case 1: $j<k$: Eq.~\eqref{eq:prob_dist} implies that
\begin{align}
\abs{\sum_{i=1}^{j-1} p_i - \sum_{i=1}^{j-1} q_i} \le 
\sum_{i=1}^{j-1} \abs{p_i - q_i}
=O(p\Delta).    
\end{align}
Then we get
\begin{align}
p_j' - q_j 
= \alpha - \sum_{i=1}^{j-1} p_i - q_j
\ge 
\alpha - \sum_{i=1}^{j-1} q_i - q_j - O(p \Delta)
\ge -O(p\Delta).
\label{eq:pjqj1}
\end{align}
Meanwhile, by Eq.~\eqref{eq:prob_dist}, we also get
\begin{align}
p_j' - q_j  \le p_j - q_j \le O(p \Delta).   
\label{eq:pjqj2}
\end{align}
Combining Eqs.~\eqref{eq:pjqj1} and \eqref{eq:pjqj2} yields
\begin{align}
\abs{p_j' - q_j } = O(p \Delta). 
\label{eq:pjqj3}
\end{align}
Furthermore, Eq.~\eqref{eq:prob_dist} also implies that
\begin{align}
\abs{\sum_{i=1}^{j} p_i - \sum_{i=1}^{j} q_i} \le 
\sum_{i=1}^{j} \abs{p_i - q_i}
=O(p\Delta).    
\end{align}
Then since $\sum_{i=1}^{j} p_i \ge \alpha$, we get $\sum_{i=1}^{j} q_i \ge \alpha - O(p\Delta)$ and hence
\begin{align}
\sum_{i=j+1}^{k-1} q_i + q'_k = O(p\Delta).    
\label{eq:qiqk}
\end{align}
Now by Eqs.~\eqref{eq:prob_dist},  \eqref{eq:cvargx}, \eqref{eq:cvargtx}, 
\eqref{eq:pjqj3} and \eqref{eq:qiqk}, we obtain
\begin{align}
\abs{\cvar{g(X)} - \cvar{g(\tilde{X})}}
&\le 
\left [\sum_{i=1}^{j-1} \abs{p_i - q_i} g(x^{(i)})
+
\abs{p_j' - q_j} g(x^{(j)}) \right .\\
&\left .\quad +
\sum_{i=j+1}^{k-1} q_i g(x^{(i)})
+ q_k' g(x^{(k)}) \right ] / \alpha\\
& \le 
\left [\sum_{i=1}^{j-1} \abs{p_i - q_i} 
+
\abs{p_j' - q_j}  \right .\\
&\left .\quad +
\sum_{i=j+1}^{k-1} q_i 
+ q_k'  \right ] (b-a)/ (2\alpha)\\
& = O(p \Delta (b-a)/\alpha).
\end{align}

\item Case 2: $j=k$: By Eqs.~\eqref{eq:prob_dist}, \eqref{eq:cvargx} and  \eqref{eq:cvargtx} we obtain
\begin{align}
\abs{\cvar{g(X)} - \cvar{g(\tilde{X})}}
&\le 
\lrb{\sum_{i=1}^{j-1} \abs{p_i - q_i} g(x^{(i)})
+ \abs{p_i' - q_i'} g(x^{(j)})}/\alpha \\
&\le 
\lrb{\sum_{i=1}^{j-1} \abs{p_i - q_i} 
+ \abs{p_i' - q_i'}}\lrb{b-a}/(2\alpha) \\
& =
\lrb{\sum_{i=1}^{j-1} \abs{p_i - q_i} 
+ \abs{\sum_{i=1}^{j-1} p_i - \sum_{i=1}^{j-1} q_i}}\lrb{b-a}/(2\alpha) \\
& \le 
\lrb{\sum_{i=1}^{j-1} \abs{p_i - q_i} }\lrb{b-a}/\alpha \\
& = O(p\Delta(b-a)/\alpha).
\end{align}

\item Case 3: $j>k$: This case can analyzed in the same way as Case 1. We only need to switch the role of $p_i$ and $q_i$, and obtain the same upper bound on $\abs{\cvar{g(\tilde{X})} - \cvar{g(X)}}$.
\end{itemize}

Combining the three cases, we know that 
\begin{align}
\abs{\cvar{g(X)} - \cvar{g(\tilde{X})}} = O(p \Delta (b-a)/\alpha)
= O(p (b-a)^2 / (\alpha M )).
\end{align}
As a result, by choosing some $M=\Theta(p(b-a)^2/(\alpha \epsilon))$,  we guarantee that 
\begin{align}
\abs{\cvar{f(X)} - \cvar{f(\tilde{X})}} =\abs{\cvar{g(X)} - \cvar{g(\tilde{X})}} \le \epsilon,
\end{align}
as claimed.

\bibliographystyle{unsrt}
\bibliography{references}

\end{document}